\documentstyle[12pt]{article}

\date{\today}

\def\be{\begin{equation}}
\def\ee{\end{equation}}
\def\bear{\begin{eqnarray}}
\def\eear{\end{eqnarray}}
\def\nn{\nonumber}

\def\half{{{1\over 2}}}

\def\wdg{{\wedge}}                              

\def\Re{{\rm Re\hskip0.1em}}
\def\Im{{\rm Im\hskip0.1em}}

\newcommand\px[1]{{\partial_{#1}}}

\newcommand\rep[1]{{{\bf {#1}}}}      
\newcommand\tr[1]{{\mbox{tr}\{{#1}\}}}          
\newcommand\com[2]{{\lbrack {#1},{#2}\rbrack}}  

\def\BZ{{{\bf Z}}}
\def\BC{{{\bf C}}}

\newcommand\MR[1]{{{\bf R}^{#1}}}               
\newcommand\MC[1]{{{\bf C}^{#1}}}               
\newcommand\MS[1]{{{\bf S}^{#1}}}               
\newcommand\MT[1]{{{\bf T}^{#1}}}               
\newcommand\CP[1]{{{\bf P}^{#1}}}              

\newcommand\SUSY[1]{{{\cal N}= {#1}}}           

\def\a{{\alpha}}
\def\b{{\beta}}

\def\lam{{\lambda}}
\def\blam{{\overline{\lambda}}}

\def\MSP{{\cal M}}                
\def\ModHyp{{\cal Y}}              
\def\ModHypZ{{\cal Y^{(0)}}}       

\def\lineb{{\cal L}} 

\def\cM{{\cal M}} 
\def\hB{{\lbrack B\rbrack}} 
\def\hF{{\lbrack F\rbrack}} 
\def\hSigma{{\lbrack \Sigma\rbrack}} 

\def\vth{{\vartheta}}
\def\bvth{{\overline{\vth}}}
\def\ba{{\overline{a}}}
\def\bz{{\overline{z}}}
\def\bw{{\overline{w}}}
\def\bu{{\overline{u}}}

\def\btau{{\overline{\tau}}}
\def\bsigma{{\overline{\sigma}}}
\def\btheta{{\overline{\theta}}}
\def\bu{{\overline{u}}}
\def\bA{{\overline{A}}}
\def\wA{{\widetilde{A}}}

\begin{document}


\begin{titlepage}
\titlepage
\rightline{hep-th/0007236}
\rightline{PUPT-1937}
\rightline{ITEP-TH-37/00}
\rightline{July 2000}
\vskip 1cm
%
%
\centerline{{\Huge Constructions of Non Commutative Instantons}}
\centerline{{\Huge on $T^4$ and $K_3$}}
\vskip 1cm
\centerline{Ori J. Ganor, Andrei Yu. Mikhailov and Natalia Saulina}
\vskip 0.5cm

\begin{center}
Department of Physics, Jadwin Hall \\
Princeton University \\
NJ 08544, USA
\end{center}

\abstract{
We generalize the spectral-curve construction of moduli spaces of
instantons on $\MT{4}$ and $K_3$ to noncommutative geometry.
We argue that the spectral-curves should be constructed inside
a twisted $\MT{4}$ or $K_3$ that is an
elliptic fibration without a section.
We demonstrate this explicitly for $T^4$ and to first order in the 
noncommutativity, for $K_3$.
Physically, 
moduli spaces of noncommutative instantons
appear as moduli spaces of theories with $\SUSY{4}$ supersymmetry in 2+1D.
The spectral curves are related to Seiberg-Witten curves
of theories with $\SUSY{2}$ in 3+1D.
In particular, we argue that the moduli space of instantons of
$U(q)$ Yang-Mills theories on a noncommutative $K_3$
is equivalent to the Coulomb branch of certain 2+1D theories
with ${\cal N} = 4$ supersymmetry.
The theories are obtained by compactifying the heterotic little-string
theory on $T^3$ with global twists. This extends a previous result
for noncommutative instantons on $\MT{4}$.
We also briefly discuss the instanton equation on generic curved
spaces.
}
\end{titlepage}
             
\tableofcontents

\section{Introduction}\label{intro}
Since a realization of noncommutative gauge theories from string
theory has been established in \cite{CDS,DH}, there has been a revived
interest in the field. In \cite{SWNCG},
noncommutative gauge theories were
obtained by studying D-branes with a large and constant
NSNS B-field (in string units) and in \cite{NekSch,Berk,SWNCG},
the moduli space of instantons on a flat noncommutative $\MR{4}$ was
studied and was shown to be a smooth space.
In later developments, the perturbative
structure of various noncommutative field theories was studied and
was shown to have a rich and surprising IR behavior 
\cite{MSV,MST,Arm} and the twistor construction
of instanton moduli spaces has been extended to various noncommutative
spaces \cite{KKO}.

In this work we will study the analytic structure of the
moduli spaces of instantons on elliptically fibered
noncommutative $\MT{4}$ and $K_3$. 
We will generalize the  ``spectral-curve'' construction for
the commutative case \cite{FMW,BJPS} to the noncommutative 
geometries.
We will see that,
as far as the holomorphic structure of the moduli space of instantons
goes, the essence of the noncommutativity can be captured
by constructing the spectral curves in a modified elliptic
fibration that does not have a section.

Aside from the theoretical interest,
the moduli spaces of instantons on noncommutative spaces
have important applications for M(atrix)-theory \cite{BFSS}.
The M(atrix)-models for the 5+1D $(2,0)$ theories are given
by quantum mechanics on moduli spaces of instantons 
on $\MR{4}$ \cite{ABKSS,WitQHB} and the M(atrix)-models of 3+1D
$\SUSY{4}$ SYM theories are derived from moduli spaces of instantons
on $\MT{4}$ \cite{GanSet}. These moduli spaces are singular but
a regularization of the moduli space of instantons was
suggested in \cite{ABS} and was later shown to be equivalent to
the moduli spaces of instantons on a noncommutative $\MR{4}$ 
\cite{NekSch,Berk}.
The special limit of $\SUSY{4}$ SYM theories requires only
the holomorphic structure of the moduli space of instantons
\cite{GanSet}, and hence our results for the moduli spaces
of noncommutative instantons are directly applicable to
the study of the M(atrix)-models of gauge theories.

Another application of moduli spaces of noncommutative instantons
is to the construction of 2+1D $\SUSY{4}$ moduli spaces of various
compactified field-theories.
Dualities in string theory and M-theory allow one to solve
the quantum-corrected moduli space of many 2+1D gauge theories with
8 supersymmetries ($\SUSY{4}$ in 2+1D) \cite{IntSei,SeiIRD}.
The argument is usually an adaptation of the reasoning of \cite{KacVaf}
as follows. The gauge theory appears in a weak coupling limit of
a certain string/M-theory background. ``Weak coupling'' means that
the Planck scale $M_p$ is infinite, in order to decouple gravity.
Duality can relate this background to another one that is strongly-coupled.
If we can use supersymmetry to argue that the coupling-constant of
the strongly-coupled system has a factor $\lambda$ that decouples from 
the moduli space in question, then we can solve the problem for
$\lambda = 0$. At this value, the system is certainly not described by
a gauge-theory any more. Nevertheless, the vacuum structure, namely
the moduli-space, is still the same.
The supersymmetry argument that one uses is the decoupling of
hyper-multiplets from vector-multiplets. If $\lambda$ is part of
a hypermultiplet and we are interested in the vector-multiplet
moduli space, we can safely argue that we can set $\lambda$ to
any value we need.

The Coulomb branch moduli space of a large class of 2+1D gauge-theories
was found in \cite{IntSei}. They studied
quiver gauge-theories \cite{DouMoo}, that is, theories with
gauge groups of the form $\prod_{i=1}^{r+1} SU(a_i N)$
(where $a_i$ are positive integers)
and bi-fundamental hypermultiplets in representations
$(\rep{\overline{a_i N}}, \rep{a_j N})$ such that the content
and the $a_i$'s are derived from an extended Dynkin diagram of
ADE type \cite{DouMoo}.
They argued that the moduli space is equivalent to
the moduli space of instantons on $\MR{4}$ with the gauge group
corresponding to the particular Dynkin diagram and $N$ is the instanton
number. Other 2+1D theories were studied in \cite{SWThreeD}.

A larger class of 2+1D questions about $\SUSY{4}$ moduli spaces
can be formulated by compactifying a 5+1D theory with (at least)
$\SUSY{(1,0)}$ supersymmetry on $\MT{3}$.
The statement of \cite{IntSei} was generalized in \cite{IntNEW}
(see also \cite{GanSet}) to the equivalence of the moduli space
of compactified little-string-theories of NS5-branes at ADE singularities
and instantons on $\MT{4}$.
A further generalization of this statement was studied in
\cite{CGKM}. There, it was argued that the moduli space of
little-string-theories compactified on $\MT{4}$ with twisted boundary
conditions is equivalent to the moduli space of instantons on
a noncommutative $\MT{4}$.
This also generalizes the result of \cite{KapSet} who showed
that the moduli space of mass deformed 2+1D $\SUSY{8}$ gauge theories
is equivalent to the moduli space of instantons on a noncommutative
$\MS{1}\times \MR{3}$.

These results give a realization of the hyper-K\"ahler moduli spaces of
instantons on $\MT{4}$ and its deformation to noncommutative instantons.
In fact, it suggests a natural definition of instanton moduli spaces
of other simply-laced groups, like $E_{6,7,8}$, on a noncommutative
$\MT{4}$ (even though the SYM theory is not defined)!

In this paper we will realize another class of hyper-K\"ahler
moduli-spaces, namely, instantons on $K_3$ and, in particular,
the deformation to instantons on a noncommutative $K_3$.
We are going to argue that the moduli space of heterotic
little-string theories \cite{SeiVBR} compactified on $\MT{3}$ with
twisted boundary conditions is equivalent to the moduli space
of instantons on a noncommutative $K_3$.
(This result was also suggested in \cite{IntCOM}.)
In particular, from special limits of this result one can obtain
the moduli space of certain 3+1D $Sp(k)$ gauge theories with matter
in the fundamental representation.

The relation between moduli-spaces of 5+1D theories that are
compactified on $\MT{3}$ and moduli spaces of instantons on 
elliptically fibered spaces ($K_3$ or $\MT{4}$) makes it clear
that the latter should have a spectral-curve type construction.
To see this, we can compactify on $\MT{2}\times\MS{1}$ and take
the limit that the radius of $\MS{1}$ is large.
In this case, we can first compactify on $\MT{2}$ and then compactify
the low-energy limit of the 3+1D theory on the large $\MS{1}$.
The low-energy limit of the 3+1D theory is constructed from
a Seiberg-Witten curve as in \cite{SWYM} and the 2-step
compactification makes it clear that the
2+1D moduli spaces are fibrations over the moduli
space of Seiberg-Witten curves (of some genus $g$),
with the fiber being the Jacobian (the $\MT{2g}$ space
of $U(1)$ gauge field configurations with zero field-strength)
of the curve \cite{SWThreeD}.
This is precisely the spectral-curve construction -- the
spectral-curve being identified with the Seiberg-Witten curve.

The paper is organized as follows.
In section (\ref{sprev}) we briefly review the spectral-curve 
constructions of holomorphic vector bundles on $\MT{4}$ and $K_3$.
In section (\ref{instf}) we extend the spectral-curve construction
of instanton moduli-spaces to the noncommutative $\MT{4}$.
We show explicitly how the spectral curves are constructed
in a twisted $\MT{4}$.
In section (\ref{instk}) we extend the construction to a 
noncommutative  $K_3$. We demonstrate explicitly, to first
order in the noncommutativity, the spectral curves.
We also discuss the instanton equations on a generic curved
manifold, to first order in the noncommutativity.
In section (\ref{lst}) we discuss the connection between noncommutative
instanton moduli spaces on $K_3$ and moduli spaces of compactified 
heterotic little-string theories.
In section (\ref{gen}) we generalize to the theories of heterotic
NS5-branes at $A_k$ singularities.
In section (\ref{swrel}) we describe the relation between 
Seiberg-Witten curves of the little-string theories compactified 
on $\MT{2}$ with twists, and the spectral-curve constructions
of noncommutative instanton moduli spaces.
We also describe the derivation of the spectral curves of
$K_3$ from T-duality.


\section{Spectral curve constructions}\label{sprev}
Holomorphic vector bundles on an eliptically fibered complex manifolds
$\cM$ with a section can be described with {\bf spectral-curves}.
We are interested in instanton moduli spaces and therefore
we will take $\cM$ to be a (real) 4-dimensional manifold.
We will follow \cite{FMW,BJPS}.

Let us denote the  base of the fibration by $B$ and let
us denote the $\MT{2}$ fiber over a generic point $p\in B$ by $F_p$.
Let $\pi: \cM\rightarrow B$ be the projection.
The idea is to first study the restriction of the bundle
to each $F_p$. A rank-$q$
holomorphic vector bundle, $V$, on $\MT{2}$ with zero first Chern class,
can be described naturally by $q$  points on $\MT{2}$.
These points then span a holomorphic curve $\Sigma\subset \cM$
which is, in general, a $q$-fold cover of $B$ with a certain
number of branch points $p_1,\dots, p_r\in B$.
Let the multiplicity of the $j^{th}$ branch-point be $n_j$.
Given $\Sigma$ we can construct the restriction of the vector
bundle to each $F_p$. This describes a vector bundle $U$
that is locally a product of $q$ line-bundles $\prod_{i=1}^q \lineb_i$
(over the local patch of $\cM$).
The structure group is locally a subgroup of $\BC^q$.
When we go around a branch point, $p_j$, we have to permute
$n_j\le q$ of the $\lineb_i$'s and this is how the nonabelian factor
which is an element of $S_{n_j}$ (the permutation group)
enters into the structure group. To describe $V$, we still
have the freedom to multiply by the pullback
$\pi^*\lineb$ of a line-bundle $\lineb$ over the base $B$.

The instanton number, $c_2(V)$, is related to the homology
class $\hSigma$ of the curve as \cite{FMW,BJPS}:
\be\label{cohkq}
\hSigma = q \hB + c_2(V)\hF,
\ee
where $\hB,\hF$ are the homology classes of $B$ and $F$, respectively.

For example, for $\cM=\MT{4}$ we can relate $c_2(V)$ to the
genus of $\Sigma$ as:
$$
2g-2 = 2 q c_2(V),
$$
while for $\cM=K_3$ we have:
$$
2g-2 = 2q(c_2(V) - q).
$$
In general, the spectral curve could have several components
and we should replace $(2g-2)$ with the sum over the distinct
components.

\section{The moduli space of instantons on $\MT{4}$}\label{instf}
Let us demonstrate the spectral curves for $\MT{4}$.
Let us, for simplicity, take $\MT{4}$ of the form $\MT{2}\times\MT{2}$
and call the first factor, the ``base'' $B$, and the second factor,
the ``fiber'' $F$.
We take $z$ to be a local coordinate on the base, such that
$$
z\sim z+1\sim z+\sigma
$$
We also take $w$ to be a local coordinate on the fiber with:
\be\label{wbc}
w\sim w+1\sim w+\tau.
\ee
Let us take the rank to be $q$ and $c_2=k$.
We restrict ourselves to vector-bundles with $c_1=0$.
The restriction of the vector-bundle to any given fiber
is a holomorphic vector bundle on $\MT{2}$ with $c_1=0$.
Let $p_0$ denote the origin of $F$.
Any line bundle with $c_1=0$ on $F$ is equivalent to a divisor
of the form $p-p_0$ with $p\in F$. The vector-bundle of rank-$q$
can be reduced to a sum of $q$ line-bundles and different
orderings are equivalent. Thus, the vector-bundle can be described
by $q$ points in $F$.
The position of the points is given locally by $q$  maps:
$$
w_a(z):B\mapsto F,\qquad a=1\dots q.
$$
The boundary conditions are:
$$
w_a(z+1) = w_{\rho(a)}(z)+n_a + m_a\tau,\,\,
(a=1\dots q),\,\,\rho\in S_q,\,\, n_a,m_a\in\BZ,
$$
and similarly for $z+\sigma$.

\subsection{Small fiber}
One can rederive the spectral curves for instantons on $\MT{4}$
by solving the instanton equations directly, in the limit that
the area of the fiber $F$ is zero.
We can take an instanton solution described by the $U(q)$ gauge fields
$A_z,A_w$ and the complex conjugate $\bA_\bz,\bA_\bw$.
The instanton equations imply:
$$
0 = F_{zw}\equiv \px{z}A_w-\px{w}A_z -i \com{A_z}{A_w}.
$$
If the fiber is very small, we can assume that $A_w$ and $A_z$ are
independent of $w$. We can then also make a gauge transformation
that ensures that $A_w$ is in a $U(1)^q\subset U(q)$ Cartan subgroup.
In the projection of the equation $F_{zw}=0$ on this Cartan subgroup,
the commutator term drops and we find that $\px{z}A_w=0$
and $\bA_\bw$ is holomorphic.
$\bA_\bw$  is a collection of $q$
periodic  variables and can be naturally identified
with $q$ points in the dual of the fiber.
The complex structure of the dual of the fiber is still $\tau$.
So $\bA_\bw$ defines a holomorphic $q$-valued map from
$B$ to the dual of
$F$ and the graph, $\Sigma$, of this map is the spectral-curve.
The dual of $F$ has the same complex structure, $\tau$.
So, as far as complex structure goes,
the instanton solution defines  a curve $\Sigma$ inside $\MT{4}$.
Locally we denote the $q$ points by:
$$
w_a(z),\,\,(a=1\dots q).
$$
The boundary conditions are that:
$$
w_a(z) = w_{\rho(a)}(z+1) = w_{\rho'(a)}(z+\sigma),
$$
for some permutations $\rho,\rho'\in S_q$.
If we set $A_z=0$, we can also calculate the instanton number as:
$$
\int \tr{F\wdg F} = \int |F_{zw}|^2 =
\sum_{a=1}^q\int |\px{z}w_a|^2 dz = \int_\Sigma d^2 w = k.
$$
Here $k$ is the number of times the curve $\Sigma$ intersects
the section $w=0$.


\subsection{Noncommutativity turned on}
We now turn on a constant bivector (contravariant 2-tensor) $\theta$
that makes $\MT{4}$ noncommutative and see how it affects
the equations.
In general, $\theta$ has 6 (real) independent components, but we will
assume that only two of them are turned on. In the complex
coordinates $z,w$, we will take it to be 
$$
\theta\equiv\theta^{zw}=-\theta^{wz},
\btheta\equiv\btheta^{\bz\bw} = (\theta^{zw})^*.
$$
Instantons on a noncommutative $\MT{4}$ have been studied in
\cite{AsNiSc} from an algebraic approach.
Noncommutative gauge theory on $\MT{d}$ has also been
studied in \cite{ASchw}-\cite{SJ}.
Below, we will describe a more direct physical approach.

The instanton equations, in the noncommutative case, imply:
\be\label{ncgineq}
\blam I = F_{\bz\bw}\equiv
   \px{\bz}A_\bw-\px{\bw}A_\bz -i A_\bz\star A_\bw +i A_\bw\star A_\bz.
\ee
Here $\lam$ is an unknown constant that, as we shall soon see,
is determined by $\theta$.
$\theta$ also enters into the definition of the $\star$-product
on the RHS.
Next, we will assume that $A_\bz = 0$ and obtain
$\blam I = \px{\bz} A_\bw$.
(We used $\px{w}A_\bz = \px{w}A_\bw = 0$.)
Let us, to start with, take the gauge group to be $U(1)$.
The last equation
implies that $A_\bw$ is given in terms of a holomorphic function
$\phi(z)$ as:
$$
A_\bw = \phi(z) + \blam \bz.
$$
The boundary conditions are:
$$
A_\bw(z,\bz) = A_\bw(z+1,\bz+1) = A_\bw(z+\sigma,\bz+\bsigma).
$$
Let us first assume, naively, that ${{\tau_2}\over {\pi i}}A_\bw$ takes
its values in a $\MT{2}$ with the same complex structure $\tau$,
in the notation of (\ref{wbc}).
We will see shortly that this assumption is not quite right,
but let us proceed with it for the time being.
Let us denote by $u$ a complex coordinate on this $\MT{2}$
with $u\sim u+1\sim u+\tau$.
Together, $(z,u)$ would define a $\MT{4}$ that is identical
in complex structure to the original $\MT{4}$.
$A_\bw$ would define a curve $\Sigma\subset\MT{4}$ by the equation:
$$
\Sigma = \left\{(z,u):
  0 = u - {{\tau_2}\over {\pi i}}A_\bw(z,\bz)\right\}.
$$
We can change the complex structure on $\MT{4}$ so
that locally holomorphic functions $\psi(z,\bz,u,\bu)$ will
be defined to satisfy:
$$
\px{\bu}\psi = (\px{\bz}+{{\tau_2}\over {\pi i}}\blam\px{u})\psi = 0.
$$
We can do this by defining:
\be\label{defuwone}
z_1 \equiv z,\qquad u_1 = u + {{\tau_2}\over {\pi i}}\blam \bz.
\ee
$$
\px{\bu_1}\equiv\px{\bu},\qquad
  \px{\bz_1}\equiv \px{\bz}+{{\tau_2}\over {\pi i}}\blam\px{u}
$$
In this complex structure, the curve $\Sigma$ that is
defined by the equation:
$$
0 = u - {{\tau_2}\over {\pi i}}A_\bw(z,\bz) 
  = u-{{\tau_2}\over {\pi i}}(\phi(z)+\blam\bz),
$$
is holomorphic.
Naively, we now face a puzzle!
If $u-{{\tau_2}\over {\pi i}}(\phi(z)+\blam\bz)$
indeed defines a curve $\Sigma$,
its cohomology class should be integral and given by an equation
like (\ref{cohkq}):
$$
\hSigma = q\hB + k\hF.
$$
But this implies that $\hSigma$ is an analytic curve in the
complex structure defined by $(u,z)$ and it cannot be also analytic
in the other complex structure defined by $(u_1,z_1)$ for $\lam\neq 0$.
In particular, if $\Sigma$ is analytic in the complex
structure defined by $u_1,z_1$, we can calculate:
$$
\int_\Sigma du\wdg dz = \int_\Sigma du_1\wdg dz_1
+{{\tau_2\blam}\over {\pi i}}\int_\Sigma dz\wdg d\bz 
= {2{\tau_2\blam}\over {\pi }} A\neq 0.
$$
where $A=\int_\Sigma \frac{dz\wdg d\bz}{2i}$ is the area of the base $B$.

The resolution of the puzzle has to do with the noncommutative
nature of the gauge group.
What are the periodicity conditions imposed on $A_\bw$?
In the commutative case we had
\be\label{permn}
{{\tau_2}\over {\pi}} A_\bw\sim {{\tau_2}\over{\pi}} A_\bw+ i(n + m \tau),
\qquad n,m\in\BZ.
\ee
where $\tau=\tau_1+i\tau_2$.
Where did this periodicity condition come from?
There are large gauge transformations $\Lambda(w,\bw)$ on the 
fiber that preserve the condition that $A_\bw$ is a constant
but shift it as follows.
For:
\be\label{gaugetr}
\Lambda = e^{{{\pi}\over {\tau_2}} \lbrack m(\tau\bw-\btau w)
            -n (w-\bw)\rbrack},
\ee
that is single-valued on $\MT{2}$, we have:
$$
A_\bw\sim \Lambda^{-1} A_\bw\Lambda +i\Lambda^{-1}\px{\bw}\Lambda
 = A_\bw + {{i\pi}\over {\tau_2}}(n+m\tau)
$$
and this implies the periodicity (\ref{permn}).
In the noncommutative case we should calculate instead:
$$
A_\bw(z,\bz)\sim \Lambda^{-1} \star A_\bw(z,\bz)\star \Lambda 
 +i\Lambda^{-1}\star\px{\bw}\Lambda
 = A_\bw\left(z-{{2\pi i\theta}\over{\tau_2}}(n+m\btau),
       \bz+{{2\pi i\btheta}\over{\tau_2}}(n+m\tau)\right) 
  +{{\pi i}\over {\tau_2}}(n+m\tau).
$$
This means  that it was incorrect to interpret the curve $\Sigma$
as a curve inside the original $\MT{4}$.
Rather, we should define a modified $\widetilde{\MT{4}}$
with the identifications:
$$
(u,z)\sim (u,z+ n' + m'\sigma)\sim
(u + n + m\tau, z-{{2\pi i\theta}\over{\tau_2}}(n+m\btau)).
$$
The curve $\Sigma$ that we defined by the equation:
$$
0 = u - {{\tau_2}\over {\pi i}}A_\bw(z,\bz) 
  = u-{{\tau_2}\over {\pi i}}(\phi(z)+\blam\bz),
$$
is uniquely defined inside $\widetilde{\MT{4}}$ and not
the original $\MT{4}$!
For a $U(q)$  gauge-group the derivation is similar.
We take the gauge transformation (\ref{gaugetr}) in a 
$U(1)$ subgroup.

Now we can determine the relation between $\lam$ and $\theta$.
If $\Sigma$ is to be holomorphic in the complex structure defined
by $(u_1,z_1)$ as in (\ref{defuwone}) then:
\be
0 = \int_\Sigma du_1\wdg dz_1  = q \int_{B'} du_1\wdg dz_1
 + k\int_{F'} du_1\wdg dz_1.
\label{new}
\ee
The base $B'$ and fiber $F'$ of $\widetilde{\MT{4}}$ are defined
from the periodicity conditions by:
\bear
B' &=& \left\{
(z=\a+\b\sigma, u = 0)
 : 0\le \a,\b\le 1
\right\},\nn\\
F' &=& \left\{
(z=-{{2\pi i\theta}\over{\tau_2}}(\a+\b\btau), u =\a+\b\tau)
 : 0\le \a,\b\le 1
\right\},\nn\\
&& \label{newbf}
\eear
Then, from (\ref{new}), and using:
\bear
\int_{B'} du_1\wdg dz_1 &=& \frac{2\blam\sigma_2\tau_2}{\pi},\nn\\
\int_{F'} du_1\wdg dz_1 &=& 
  \int_{F'} du\wdg dz -\frac{\blam\tau_2}{i\pi} \int_{F'} dz\wdg d\bz
   =  -4\pi \theta -{8\pi \blam |\theta|^2}.
\nn
\eear
one finds: 
\be\label{lamkq}
\blam = 
 {{2\pi^2 k \theta}\over {q\sigma_2\tau_2 -4\pi^2 k|\theta|^2}}.
\ee
This agrees with the formula in \cite{AsNiSc}.

\subsection{Irreducibility of the curves}\label{irred}
The moduli spaces of commutative instantons have singularities.
On $\MR{4}$ the singularities correspond to instantons of zero size.
On $\MT{4}$, the singular instanton configurations correspond
to reducible spectral curves.
We have seen that to describe a commutative instanton we need to
find a holomorphic curve in the class:
$$
\hSigma = q\hB + k\hF.
$$
A reducible instanton configuration will be of the form
$\Sigma_1\cup\Sigma_2$ where:
\be\label{sigsig}
\hSigma_1 = q_1\hB+k_1\hF,\,\,
\hSigma_2 = q_2\hB+k_2\hF,\qquad
q=q_1+q_2,\,\, k=k_1+k_2.
\ee
It can be thought of as an embedding of the instantons in a 
$U(q_1)\times U(q_2)\subset U(q)$ subgroup such that
the $U(q_1)$ instantons have instanton number $k_1$.
One of the advantages of turning on the noncommutativity is
that the moduli spaces can become nonsingular.
For $\MR{4}$ this has been shown in \cite{NekSch,Berk}.

In the case of $\MT{4}$ we can show that the noncommutative
spectral-curves are always irreducible if the greatest
common divisor, $gcd(k,q)=1$.
First note that we cannot split the $k$ instantons in two
by embedding $k_1$ instantons in $U(q_1)$ and the other
$k_2$ in $U(q_2)$. This is because $\lam$ in (\ref{ncgineq})
depends on $k$ (see (\ref{lamkq})) and unless $k_1/q_1=k_2/q_2$
we will have to have different $\lam$'s for $U(q_1)$ and $U(q_2)$
which is prohibited. In general,
suppose that a certain spectral curve is reducible
as in (\ref{sigsig}).
Both $\Sigma_1$ and $\Sigma_2$ would have to be holomorphic
in the same complex structure as $\Sigma$.
If $gcd(k,q)=1$ this would imply that $\hF$ should be also
analytic  (in $H^{(1,1)}$) for that particular complex structure.
But as we have seen, for generic $\theta$ this is not the case,
so the curves are always irreducible for $gcd(k,q)=1$.

The moduli-space itself is a $\MT{2kq}$-fibration (the dual
of the Jacobian) over the moduli space of spectral-curves.
Although the fibers always correspond to irreducible curves
they can become degenerate when cycles shrink to zero size.

\subsection{The spectral bundle}\label{spbund}
So far we have set $A_\bz=0$.
Let us set $A_\bz\neq 0$ but keep $A_\bz$ a function of $(z,\bz)$
alone.
The third instanton equation implies:
$$
\rho I =\px{z} A_\bz -\px{\bz}A_z
  + \px{w}A_\bw - \px{\bw}A_w
  + i A_z\star A_\bz - i A_\bz\star A_z
  + i A_w\star A_\bw - i A_\bw\star A_w
 = \px{z} A_\bz -\px{\bz}A_z.
$$
Here we used the fact that $A_\bw$ is independent of $w$.
As in the commutative case, it can be seen that $\rho=0$
(for $c_1 = 0$) and that $A_\bw$ can be lifted to a $U(1)$
(commutative) flat connection over the spectral curve $\Sigma$.

The spectral bundle is described by a flat connection
on the curve $\Sigma$, of genus $g$ (which is generically $kq+1$).
It is therefore specified by specifying a point in the 
dual of the Jacobian of the curve. The Jacobian is $\MT{2g}$ and
a point on this $\MT{2g}$ corresponds to a map from 
$\pi_1(\Sigma)\rightarrow U(1)$, that describes the $U(1)$ Wilson lines
around the 1-cycles of $\Sigma$.
The holomorphic structure of the Jacobian varies when the curve is varied.
The moduli space of instantons is thus, locally, a $\MT{2g}$
fibration over the moduli space of $\Sigma$'s.

Let us also note that the embedding
$\rho:\Sigma\rightarrow\widetilde{\MT{4}}$
induces a map $\rho^*:\pi_1(\Sigma)\rightarrow
 \pi_1(\widetilde{\MT{4}})$.
Let us take $\tilde{\gamma}_1,\tilde{\gamma}_2$ to 
be generators of $\pi_1(F')\subset \pi_1(\widetilde{\MT{4}})$,
i.e. corresponding to 1-cycles on the fiber, $F'$.
Let us take $\tilde{\gamma}_3,\tilde{\gamma}_4$ to 
be generators of $\pi_1(B')\subset \pi_1(\widetilde{\MT{4}})$,
i.e. corresponding to 1-cycles on the base, $B'$.
Then for $i=1,2$ $\rho^*(\gamma_i) = q\tilde{\gamma}_i$,
where $\gamma_i$ are generators of $\pi_1(\Sigma)$.
For $i=3,4$ we have similarly, $\rho^*(\gamma_i) = k\tilde{\gamma}_i$.
The $U(1)$ holonomies along 
the cycles corresponding to  $\gamma_1,\dots,\gamma_4$, generate
a $\MT{4}$ subset of the Jacobian $\MT{2g}$.
This $\MT{4}$ is actually the dual, $\MT{4}^\vee$, of the
original $\MT{4}$.
It is not hard to see that as the curve $\Sigma$ varies
the dual of the Jacobian is locally of the form
$\MT{4}^\vee\times \MT{2g-4}$. Here $\MT{4}^\vee$ is the dual of the
original $\MT{4}$. (In subsection (\ref{examp}) we will see an
explicit example.)
The complex structure of $\MT{2g-4}$ varies when $\Sigma$
varies but the complex structure of the $\MT{4}^\vee$ remains fixed.
Globally, the moduli space is a $\MT{4}^\vee$ fiber-bundle.
The fiber is this fixed $\MT{4}^\vee$ that corresponds
to the $U(1)$ holonomies along $\gamma_1,\dots,\gamma_4$.
The structure group of this fiber-bundle is 
$\BZ_q^2\otimes\BZ_k^2$, because shifting the holonomy along
$\gamma_1$, say, by ${{2\pi}\over {k}}$, does not change
the induced holonomy along $\tilde{\gamma}_1$.

\subsection{Explicit formulas for the curves}
We can write down explicit formulas for the curves using 
$\Theta$-functions on $\MT{4}$. (See \cite{GriHar} for more details.)
Let us choose real coordinates $x_1,\dots,x_4$ such that
$0\le x_i\le 1$ and the periodic boundary conditions
are given by $x_i\sim x_i+1$.
Let us denote by $e_i$ the path from $x_i=0$ to $x_i=1$
keeping the other $x_j$'s equal to zero.
Consider the cohomology class:
$$
\psi = m_1 dx_1\wdg dx_3 + m_2 dx_2\wdg dx_4.
$$
where $k=m_1$ and $q=m_2$ are positive integers.
We wish to find explicit formulas for holomorphic curves in
a homology class $\hSigma$ that is Poincare dual to $\psi$.

A $\MT{4}$ has a $\CP{1}$-worth of complex structures.
Given a particular complex structure, one can describe it 
by picking complex coordinates
$z_1,z_2$ that are linear combinations
of the $x_i$'s. Of course, $z_1,z_2$ can be chosen up to 
a $GL(2,\BC)$ transformation.
It can be shown (see \cite{GriHar})
that if there exists a holomorphic 
curve of class $\hSigma$ then $z_1,z_2$ can be chosen such that
the $2\times 4$ periods are of the form:
\bear
\int_{e_i} dz_\a &=&
\delta_{i\a} m_i,\qquad i=1,2,\qquad \a=1,2,\nn\\
\int_{e_{i+2}} dz_\a &=& Z_{i\a},\qquad
 i=1,2\qquad \a=1,2,
\nn
\eear
and the  $2\times 2$ matrix $Z$ is symmetric and its imaginary part,
$\Im Z$, is a positive matrix.

Now define $m_1 m_2 = k q$ $\theta$-functions as follows.
For $0\le l_1 < m_1$ and $0\le l_2 < m_2$ we set:
$$
\Theta_{l_1 l_2}(z_1, z_2) \equiv
\sum_{N_1, N_2\in\BZ}
e^{\pi i \sum_{\a,\b=1,2} Z_{\a\b} N_\a (N_\b + 2 m_\b^{-1} l_\b)
+ 2\pi i \sum_{\a=1,2} (N_\a + m_\a^{-1} l_\a) z_\a}.
$$
Let us look at an equation of the form:
\be\label{thetaq}
0 = 
\sum_{l_1 = 0}^{m_1-1}
\sum_{l_2 = 0}^{m_2-1} A_{l_1 l_2}\Theta_{l_1 l_2}(z_1, z_2)
\ee
It depends on the $m_1 m_2 = k q$ complex parameters $A_{l_1 l_2}$.
It can be shown to describe a curve of class $\hSigma$.
The $\Theta$-functions are linearly independent.
Since the overall factor is unimportant, we see that the moduli space
of such curves is equivalent to $\CP{kq-1}$.
The moduli space of curves of class $\hSigma$  is of dimension $kq+1$.
The missing parameters can be described as translations of the
curve as a whole.
The translations are described by two parameters, one for
$z_1\rightarrow z_1+\epsilon$ and the other for
$z_2\rightarrow z_2+\epsilon$.

\subsection{Examples: Instanton number $k=1,2$}\label{examp}
As a simple example, let us take $q=1$ and $k=1$.
The spectral curves are of genus $g=2$ and there is only
one such curve inside $\widetilde{\MT{4}}$, up to translations.
Thus the moduli space of spectral curves is $\widetilde{\MT{4}}$.
The dual of the Jacobian of the curve is a fixed $\MT{4}_J$.
According to the discussion at the end of subsection (\ref{spbund}),
this $\MT{4}_J$ is the dual $\widetilde{\MT{4}}^\vee$
 and the overall
moduli-space is of the form
$\widetilde{\MT{4}}\times\widetilde{\MT{4}}^\vee$.

As a second example, let us take $q=1$ and $k=2$, that is,
two $U(1)$ instantons.
The commutative moduli space would be formally the space
of point-like instantons, $(\MT{4})^2/\BZ_2$ multiplied by
an extra $\MT{4}$ factor corresponding to the overall $U(1)$
holonomies. The overall $\MT{4}$ factor corresponds to
the 4 overall degrees of freedom of translating  the spectral
curve as a whole.

 From the spectral-curve construction, we obtain spectral curves
that are of genus $k q +1 = 3$. Their Jacobian is a varying $\MT{6}$.
As we have seen in subsection (\ref{spbund}),
we can separate the $\MT{6}$ into a locally fixed $\MT{4}$ and a varying
$\MT{2}$.
The moduli-space of spectral curves
is $\CP{1}$. and we therefore obtain a $\MT{2}$ fibration
over $\CP{1}$ which must correspond to a $K_3$.
If we add the extra $\MT{4}^\vee$
factor that we separated from the dual of the Jacobian
we get a moduli space that is a $\MT{4}^\vee$ fiber-bundle over
$K_3$, with the structure group being $\BZ_2$ (corresponding
to shifts by half lattice-vectors in the fiber $\MT{4}$).

We can be more precise and calculate the number of singular fibers
over this $\CP{1}$.  From (\ref{thetaq}) we see that the curves
inside $\MT{4}$ are given by an equation of the form
$0=\Theta_{00} - \lambda\Theta_{10}$,
where $\Theta_{00}$ and $\Theta_{10}$
are two sections of the same line-bundle over $\MT{4}$ and 
$\lambda\in\CP{1}$ is a coordinate on the moduli space of
spectral-curves. For a generic point in $\MT{4}$, 
the values of $\lbrack\Theta_{00},\Theta_{10}\rbrack$ define a unique
point in $\CP{1}$. However, there are 4 points inside $\MT{4}$
where $\Theta_{00}=\Theta_{10}=0$. (The number $4$
is obtained by calculating
the intersection number of the two curves 
$\Theta_{00}=0$ and $\Theta_{10}=0$.)
These points do not define a unique point in $\CP{1}$, but 
in the manifold $\MT{4}$ with 4 points blown up, every point corresponds
to a unique point on $\CP{1}$. Thus, generically,
this manifold is a fibration
of the genus-$3$ spectral-curves over the base $\CP{1}$.
The Euler number of $\MT{4}$ with 4 blown up points is $\chi=4$.
On the other hand, the Euler number of a genus-$3$ Riemann surface
fibered over $\CP{1}$ is $\chi(\CP{1})\times\chi(\Sigma_3) = -8$.
The mismatch is accounted for by noting that there must be $d$
points on the base $\CP{1}$ where a cycle of the spectral-curve
shrinks. This would bring the Euler number up to $\chi=-8+d$
and for $d=12$ we get a match.

For $\lbrack\lambda_1,\lambda_2\rbrack\in\CP{1}$, the curve:
$$
\lambda_1 \Theta_{00}(z_1,z_2) + \lambda_2 \Theta_{10}(z_1,z_2) = 0,
$$
is singular when:
$$
\lambda_1 \px{z_i}\Theta_{00}(z_1,z_2) 
  +\lambda_2 \px{z_i}\Theta_{10}(z_1,z_2) = 0,\qquad i=1,2.
$$
Eliminating the $\lambda_i$'s, this gives us two equations:
$$
\Phi_i(z_1,z_2)\equiv \Theta_{10} \px{z_i}\Theta_{00}
  -\Theta_{00} \px{z_i}\Theta_{10} = 0,\qquad i=1,2.
$$
Note that if $\Theta_{00}$ and $\Theta_{10}$ are sections
of the same line-bundle $\lineb$, then $\Phi_i$ are sections of
$\lineb^2$. The divisor of zeroes for each $\Phi_i$ is 
therefore $2\hB+4\hF$. Thus the number of solutions to
$\Phi_1 = \Phi_2 = 0$ is 16.
Out of these, 4 solutions correspond to $\Theta_{00}=\Theta_{10}=0$.
These are not singular points, since the derivatives are in general
nonzero.
Thus we are left with 12 singular points.

For $m_1=2$ and $m_2=1$ we have:
\bear
\Theta_{00}(z_1, z_2) &=&
\sum_{N_1,N_2\in\BZ}
  e^{\pi i ( Z_{11} N_1^2 + 2 Z_{12} N_1 N_2 + Z_{22} N_2^2
   + 2 N_1 z_1 + 2 N_2 z_2)},\nn\\
\Theta_{10}(z_1, z_2) &=&
\sum_{N_1,N_2\in\BZ}
  e^{\pi i ( Z_{11} N_1^2 + 2 Z_{12} N_1 N_2 + Z_{22} N_2^2
   + 2 N_1 z_1 + 2 N_2 z_2 + Z_{11} N_1 + Z_{12} N_2 +z_1)},\nn
\eear
The $\MT{4}$ is generated by the columns of the matrix:
\be\label{ZZ}
\left(\begin{array}{cccc}
1 & 0 & \frac{1}{2}Z_{11} & \frac{1}{2}Z_{21} \\
0 & 1 & Z_{12} & Z_{22} \\
\end{array}\right),\qquad Z_{21} = Z_{12}.
\ee
%

Now let us describe the Jacobian of the $g=3$ curve.
The Jacobian is a $\MT{6}$ and to describe it, we need to pick
3 holomorphic 1-forms, $\psi_i$ ($i=1,2,3$) on $\Sigma$ and
Integrate them over the 6 1-cycle generators of $H_1(\Sigma)$.
This will give us a basis for a lattice in $\MC{3}$.
Let $\rho:\Sigma\subset \MT{4}$ be the embedding of the curve.
We can take $\psi_i=\rho^* dz_i$ ($i=1,2$).
Let $C'_a$ ($a=1\dots 6$) be a$a=1\dots 4$) 
be a basis of 1-cycles of $\Sigma$ and
let $C_a$ ($a=1\dots 4$) be a basis of 1-cycles of $\MT{4}$
corresponding to the columns of (\ref{ZZ}).
We can choose $C'_a$ such that:
$$
\rho(C'_1)=C_1,\,\,
\rho(C'_2)=\rho(C'_3)=C_2,\,\,
\rho(C'_4)=C_3,\,\,
\rho(C'_5)=\rho(C'_6)=C_4.
$$
We can now choose $\psi_3\in H^{(1,0)}(\Sigma)$ such that:
$$
\int_{C'_1}\psi_3 = \int_{C'_2}\psi_3 = 0.
$$
The period matrix for $\Sigma$ (whose elements are 
the integrals $\int_{C'_a}\psi_i$) is of the form:
$$
\left(\begin{array}{cccccc}
1 & 0 & 0 & \frac{1}{2}Z_{11} & \frac{1}{2}Z_{21} & \frac{1}{2}Z_{21} \\
0 & 1 & 1 & Z_{12} & Z_{22} & Z_{22} \\
0 & 0 & 1 & x & y & z \\
\end{array}\right).
$$
After a linear transformation this becomes:
$$
\left(\begin{array}{cccccc}
1 & 0 & 0 & \frac{1}{2}Z_{11} & \frac{1}{2}Z_{12} & \frac{1}{2}Z_{12} \\
0 & 1 & 0 & Z_{12}-x & Z_{22}-y & Z_{22}-z \\
0 & 0 & 1 & x & y & z \\
\end{array}\right).
$$
The condition that $\Sigma$ is analytic implies that the second
$3\times 3$ block of this matrix should be symmetric (see \cite{GriHar}).
It follows that $x=\frac{1}{2}Z_{12}$ and $y=Z_{22}-z$.
A holomorphic change of basis: $\psi'_1 = \psi_1$,
$\psi'_2 = \psi_2+\psi_3$ and $\psi'_3 = \psi_3-\psi_2$ and
an $SL(6,\BZ)$ transformation renders the Jacobian in the form of
a $\MT{6}$ generated by the columns of:
$$
\left(\begin{array}{cccccc}
1 & 0 & \frac{1}{2}Z_{11}& Z_{12} & 0 & 0 \\
0 & 1 & Z_{12}& 2 Z_{22} & 0 & 0 \\
0 & 0 & \frac{1}{2}Z_{12}   & Z_{22} & w & 1 \\
\end{array}\right),
$$
Here $w=Z_{22}-2z$.
We actually need the dual of that $\MT{6}$.
It is generated by:
$$
\left(\begin{array}{c}
1 + {{\Re Z_{12}\Im Z_{12}-\Re Z_{11}\Im Z_{22}}
  \over {\Im Z_{11}\Im Z_{22} -\Im Z_{12}\Im Z_{21}}}i \\
{{\Im Z_{12}\Re Z_{11}-\Im Z_{11}\Re Z_{12}}
  \over {2(\Im Z_{11}\Im Z_{22} -\Im Z_{12}\Im Z_{21})}}i \\
0  \\
\end{array}\right),\,
\left(\begin{array}{c}
 {{2(\Re Z_{22}\Im Z_{21}-\Re Z_{12}\Im Z_{22})}
    \over {\Im Z_{11}\Im Z_{22} -\Im Z_{12}\Im Z_{21}}}i \\
  1+{{\Im Z_{12}\Re Z_{21}-\Im Z_{11}\Re Z_{22}}
    \over {\Im Z_{11}\Im Z_{22} -\Im Z_{12}\Im Z_{21}}}i \\
0 \\
\end{array}\right),\,
\left(\begin{array}{c}
 {{\Im Z_{22}}
    \over {\Im Z_{11}\Im Z_{22} -\Im Z_{12}\Im Z_{21}}}i \\
-{{\Im Z_{12}}
  \over {2(\Im Z_{11}\Im Z_{22} -\Im Z_{12}\Im Z_{21})}}i \\
0 \\
\end{array}\right)
$$
and
$$
\left(\begin{array}{c}
-{{\Im Z_{21}}
    \over {\Im Z_{11}\Im Z_{22} -\Im Z_{12}\Im Z_{21}}}i \\
{{\Im Z_{11}}
  \over {2(\Im Z_{11}\Im Z_{22} -\Im Z_{12}\Im Z_{21})}}i \\
0 \\
\end{array}\right),\,
\left(\begin{array}{c}
0 \\
0 \\
1 \\
\end{array}\right),\,
\left(\begin{array}{c}
-{{\Re Z_{22}\Im Z_{12}-\Re Z_{12}\Im Z_{22}}
  \over {\Im Z_{11}\Im Z_{22} -\Im Z_{12}\Im Z_{21}}}i \\
-\frac{1}{2}
-{{\Re Z_{12}\Im Z_{12}-\Re Z_{22}\Im Z_{11}}
  \over {2(\Im Z_{11}\Im Z_{22} -\Im Z_{12}\Im Z_{21})}}i \\
w  \\
\end{array}\right).
$$
This describes the dual $(\MT{4})^\vee$ fibered over the $\MT{2}$
with (variable) complex structure $w$. The fibration is described
by a translation by half a lattice when going around one of the cycles
of the $\MT{2}$.

\def\wstar{{\tilde{star}}}
\def\wA{{\widetilde{A}}}

\section{Instantons on $K_3$}\label{instk}
After having analyzed instantons on $\MT{4}$, we move on to
noncommutative instantons on other complex manifolds where the
metric is not flat.
As in the case of $\MT{4}$, we wish to study the holomorphic
structure of the moduli space of instantons.

As a first step, we have to discuss the form of the instanton
equations on the noncommutative manifold.
As in the $\MR{4}$  case, the instanton equations can be written
down once one defines an associative $\star$-product on functions
and on 1-forms. The constant bivector (antisymmetric
contravariant tensor) $\theta^{ij}$ is now replaced
with a covariantly constant bivector that we still denote by
$\theta^{ij}$.
The construction of the $\star$-product can be found in
\cite{Konts,Fed,CatFel}. We will only need the lowest order terms
which we will review shortly.

We will also restrict ourselves to bivectors, $\theta^{ij}$,
such that the inverse matrix $(\theta^{-1})_{ij}$ is
a sum of holomorphic and anti-holomorphic 2-forms.
The only 4-dimensional compact
manifolds with a covariantly constant holomorphic 2-form
are $\MT{4}$ and $K_3$, so our next example is a $K_3$ surface.
Furthermore, we will consider the special case of elliptically
fibered $K_3$'s with a section.

We will first review the constructions of commutative
instantons on elliptically fibered $K_3$'s and the construction
of spectral curves. We will then discuss the instanton equations
on a curved noncommutative space  and we will calculate the 
corrections to the spectral curve construction for a $K_3$.
Finally, we will compare with predictions from T-duality
in section (\ref{swrel}).

We will begin with some preliminaries from geometry.

\subsection{The geometry of elliptically fibered $K_3$}\label{geomef}
Let us start by reviewing the geometry
and the construction of holomorphic curves inside an
elliptically fibered $K_3$.
We can parameterize the $K_3$ as:
$$
y^2 = x^3 - f_8(z) x - g_{12}(z),
$$
where $z$ is a coordinate on the base, $\BC$, and $(x,y)$ are coordinates
on $\MC{2}$. The base is made into $\CP{1}$ by adding the point
$z=\infty$ and defining the good coordinates near $z=\infty$ as
follows:
$$
w={1\over z},\qquad \xi = {x\over {z^4}},\qquad 
\eta = {y\over {z^6}}.
$$
This means that there is a line-bundle $\lineb={\cal O}(2)$ on $\CP{1}$
and $x$ is a section of $\lineb^2$ and $y$ is a section of $\lineb^4$.

We denote the homology class of the fiber by $\hF$ and the homology
class of the section $x=y=\infty$ by $\hB$.
We have the intersections:
$$
\hB\cdot \hB = -2,\,\,
\hF\cdot \hF = 0,\,\,
\hB\cdot \hF = 1.
$$
For any analytic curve $\Sigma\in K_3$, we have the adjunction formula:
$$
2 g(\Sigma)-2 = \Sigma\cdot\Sigma.
$$
An expression of the form \cite{FMW}:
$$
\sum_{k=0}^n a_k(z) x^k + y\sum_{k=0}^{m} b_k(z) x^k = 0,
$$
defines an analytic curve.
Here we take $a_k(z)$ to be a section of $\lineb_0\otimes\lineb^{-2k}$
and $b_k(z)$ to be a section of $\lineb_0\otimes\lineb^{-3-2k}$.
We take $\lineb_0={\cal O}(q+l)$ with $q$ being the largest of
$3+2m$ and $2n$.
If $2m+3> 2n$,
then $q=2m+3$ and $a_k$ is of degree $2m-2k+l+3$ and $b_k$
is of degree $2m-2k+l$.
Let us count the number of solutions
to $x=y=\infty$.
The leading term behaves like  
$P_{2m-2n+l+3}(z) x^n  + Q_{l}(z) y x^m = 0$,
where $P_{2m-2n+l+3}$ and $Q_l$ are polynomials of degrees
$(2m-2n+l+3)$ and $l$ respectively.
If $x=\infty$ we need $y\sim x^{3/2}$.
Near $x=y=\infty$, the good coordinate
is $\rho\equiv x^{-1/2}\sim {x\over y}$.
For $2m+3>2n$ the equation becomes:
$P_{2m-2n+l+3}(z) \rho^{2m+3-2n}  + Q_{l}(z) = 0$.
When we set $\rho=0$ we have to solve $Q_l(z)=0$.
In general, $Q_l(z)$ will have $l$ zeroes
and this is the number of solutions of $x=y=\infty$.
We equate this with $c_2-2q$ and find $c_2=2q+l$.
If $2m+3<2n$ we find in a similar fashion that $q=2n$.
Then $a_k$ is of degree $2n-2k+l$ and $b_k$
is of degree $2n-2k-3+l$.
we have to solve  
$P_{l}(z) x^n  + Q_{2n-2m-3+l}(z) y x^m = 0$
for $x=y=\infty$.
We find that there are $l$ solutions, as before,
and therefore $c_2 = 2q+l$ again.
The homology class of $\Sigma$ is therefore:
$$
\hSigma = q\hB+(2q+l)\hF.
$$
Its genus is given by:
$$
2 g(\Sigma)-2 = \Sigma\cdot\Sigma = 2q(q+l).
$$

When the curves are spectral curves that describe instantons of $U(q)$
at instanton number $k$, we identify $q=F\cdot\Sigma$ and
$k \equiv c_2 = 2q+l$.

Later, it will be more convenient to change coordinates
locally to $w$ and $a$ that we define
as follows.
We take $w$ to be a local coordinate on the fiber and $a(z)$
to be a local coordinate on the base.
Locally, and away from the singular fibers,
we can fix a basis $\a,\b$ of $H^1(F)$,
the 1-cycles of the $\MT{2}$ fiber.
We take $w$ to be a coordinate on the fiber, with the identifications:
$$
w\sim w+n+m\tau(z),\qquad n,m\in\BZ
$$
where $\tau(z)$ is the complex structure of the fiber
(defined locally up to $SL(2,\BZ)$).
$w$ is chosen so that the integrations on the fiber give:
$$
\oint_\a dw = 1,\qquad \oint_\b dw = \tau(z).
$$
We then define $a(z)$ locally, as in
the Seiberg-Witten theory, by the formula:
\be\label{adef}
{{da}\over {dz}} = \oint_{\a} {{dx}\over y},
\ee
We will also need the integral on the other cycle, $a_D(z)$:
\be\label{bdef}
{{da_D}\over {dz}} = \oint_{\b} {{dx}\over y}.
\ee
We now write the complex structure of the fiber as
$\tau(a) = \tau_1 + i\tau_2$.
The metric is given by:
$$
ds^2 = \tau_2 da d\ba 
      +{\rho\over {\tau_2}} \left| 
  dw + i {{\tau'}\over {2\tau_2}}(w-\bw) da\right|^2.
$$
Here $\rho$ is the area of the fiber and we set it to $\rho=1$,
for simplicity.

\subsection{Small fiber}
As in the case of $\MT{4}$,
we can solve the instanton equations directly
in the limit that the area of the fiber shrinks to zero.
In this subsection we will review the commutative
$U(q)$ theories \cite{FMW} and in the next subsection we will discuss 
the noncommutative case.
The instanton equation implies:
$$
0 = F_{\bz\bw}\equiv \px{\bz}A_\bw-\px{\bw}A_\bz -i \com{A_\bz}{A_\bw},
$$
Now we assume that the fiber $\MT{2}$ is small and therefore
$A_\bw$ is independent of $w$ and $\bw$.
However, 
we should not conclude that $A_\bz$ is independent of $w$ because
it is not periodic in $w$.
Rather, under:
$$
w\rightarrow w +\tau(z),
$$
we have:
$$
A_\bz\rightarrow A_\bz -\btau'(\bz) A_\bw.
$$
Therefore:
\be\label{abzper}
A_\bz +{{\btau'(\bz)}\over {\tau(z)-\btau(\bz)}}(w-\bw) A_\bw,
\ee
is periodic in $w$ and we can assume that it is independent of $w$.
$$
\px{\bz}A_\bw -\px{\bw}A_\bz = 
\px{\bz}A_\bw -{{\btau'(\bz)}\over {\tau(z)-\btau(\bz)}} A_\bw
 = {1\over {\tau_2}} \px{\bz}\left(\tau_2 A_\bw\right).
$$
We can choose $A_\bw$ to be a diagonal $q\times q$ matrix.
The instanton equation then becomes:
$$
\px{\bz}\left(\tau_2 A_\bw\right) = 0.
$$
\be\label{phidef}
A_\bw ={{\pi}\over {\tau_2}}\phi(z),
\ee
where $\phi(z)$ is analytic in $z$.
The Wilson lines along $\a,\b$ are related to $A_\bw$ as:
$$
\int_\a A = 2\Im A_\bw,\qquad
\int_\b A = 2\Im (\btau A_\bw).
$$
$\tau_2 A_\bw$ has the following periodicity:
$$
{{\tau_2}\over {\pi}} A_\bw\sim {{\tau_2}\over {\pi}} A_\bw +n + m\tau(z),
\qquad n,m\in \BZ.
$$
Locally, $\phi(z)$ defines $q$ holomorphic functions
from $B$ with values in the fiber and they fit together to describe
a holomorphic curve $\Sigma\subset K_3$.

\subsection{Instanton equations on a noncommutative manifold}
To define the instanton equations on a curved manifold
we need to modify the $\star$-product.
We assume that $\theta^{ij}$ is covariantly constant.
(Throughout this section,
we will use small letters $a,b,c,d,e,\dots,i,j,\dots$ for 
tensor indices.)
The $\star$-product is an associative product on functions
and tensors that to lowest order looks like:
\bear
T_{a_1\dots a_p}\star S_{b_1\dots b_q} &=&
   T_{a_1\dots a_p} S_{b_1\dots b_q}
  + i\theta^{cd}\nabla_c T_{a_1\dots a_p}
      \nabla_d S_{b_1\dots b_q}+ O(\theta)^2,\nn
\eear
Here $\nabla$ is the covariant derivative:
\bear
\nabla_c  T_{a_1\dots a_p} &\equiv&
  \px{c}T_{a_1\dots a_p} -\Gamma^{b}_{c a_1} T_{b a_2\dots a_p}
  - \cdots -\Gamma^{b}_{c a_p} T_{a_1\dots a_{p-1} b}.
\nn
\eear
where $\Gamma_{ab}^c$ is the Christofel symbol.
Associativity is required to determine the higher order corrections.
The complete expression can be found in \cite{Konts,Fed,WilCom,CatFel}.

On a curved manifold, the covariant derivative is not a derivation
for the $\star$-product, i.e. 
$$
\nabla_c(T_{a_1\dots a_p}\star S_{b_1\dots b_q}) \neq
\nabla_c T_{a_1\dots a_p}\star S_{b_1\dots b_q}
 +T_{a_1\dots a_p}\star \nabla_c S_{b_1\dots b_q}.
$$
For example, it is easy to check that for a scalar and a vector:
\bear
\nabla_b(\phi\star A_a) 
  -\nabla_b\phi\star A_a -\phi\star \nabla_b A_a
 &=&
i\theta^{cd}\nabla_c\phi (\nabla_b\nabla_d-\nabla_d\nabla_b)A_a
 + O(\theta)^2
\nn\\ &=&
i\theta^{cd}\nabla_c\phi {R_{bda}}^e A_e + O(\theta)^2.
\nn
\eear
Here ${R_{abc}}^d$ is the curvature tensor.

Now suppose we have a gauge field $A_a$. We can attempt to
define the field-strength as:
$$
F_{ab} = \nabla_a A_b-\nabla_b A_a -i A_a\star A_b + i A_b\star A_a.
$$
The instanton equations will then be:
$$
F_{ab} - \sqrt{g}\epsilon_{abcd} F^{cd} = \lambda'(\theta^{-1})_{ab},
$$
where $\lambda'$ is a constant.
However, these equations will not be invariant under a gauge
transformation:
$$
A_a\rightarrow i\Lambda^{-1}\star\nabla_a\Lambda
 + \Lambda^{-1} \star A_a\star \Lambda.
$$
To first order we calculate:
\bear
F_{ab} &=& \nabla_a A_b-\nabla_b A_a -i A_a A_b + i A_b A_a
 +\theta^{kl}\nabla_k A_a\nabla_l A_b
 -\theta^{kl}\nabla_k A_b\nabla_l A_a + O(\theta)^2
\nn
\eear
For a $U(1)$ gauge field we then find:
\bear
F_{ab}' - \Lambda^{-1} \star F_{ab}\star \Lambda &=&
\theta^{kl}{R_{abl}}^m\lbrack
(\Lambda^{-1}\nabla_m\Lambda)(\Lambda^{-1}\nabla_k\Lambda)
-2iA_m \Lambda^{-1}\nabla_k\Lambda\rbrack + O(\theta)^2.
\label{curdep}
\eear
(We have used the identity
${R_{alb}}^m - {R_{bla}}^m = {R_{abl}}^m$.)
In general, this term is nonzero.
Note that for a Hyper-K\"ahler 4-manifold, ${R_{abl}}^m$ is self-dual
in the $a,b$ indices.
The anti-self-dual part of the field-strength
is therefore gauge invariant for Hyper-K\"ahler manifold.

\subsection{Spectral curves for a noncommutative $K_3$}
We now specialize to $K_3$.
The noncommutativity is specified by a symplectic form that is 
a bivector (an antisymmetric
contravariant tensor of rank 2) $\theta^{ij}$.
We will denote by $(\theta^{-1})_{ij}$ the inverse 2-form
such that $\theta^{ij}(\theta^{-1})_{jk} = \delta^i_k$.
We also assume that $\theta^{-1}$ is a sum of a $(2,0)$ and $(0,2)$ forms.
Let $\omega$ be a covariantly constant $(2,0)$ form on $K_3$.
In terms of the local holomorphic
coordinates $a$ and $w$, defined near (\ref{adef}),
we have locally:
$$
\omega = da\wdg dw = dz\wdg {{dx}\over y}.
$$
We take the $(2,0)$ part of $\theta$ to be proportional
to $\omega$ so that:
$$
\theta^{-1} = \vth^{-1} da\wdg dw + \bvth^{-1}\ba\wdg d\bw.
$$
Here $\vth$ is a proportionality constant and $\bvth$ is
its complex conjugate.

We now shrink the fiber to zero and take the metric to be:
$$
ds^2 = \tau_2 da d\ba
      +{\rho\over {\tau_2}} \left|
  dw + i {{\tau'}\over {2\tau_2}}(w-\bw) da\right|^2.
$$
Here:
$$
\tau_2 \equiv {i\over 2}(\btau(\ba)-\tau(a)),\qquad \tau'\equiv \tau'(a),
$$
and $\rho\rightarrow 0$ is the area of the fiber.
As in (\ref{abzper}), the boundary conditions on $A_a$ imply,
for a small fiber, that:
$$
A_a(w,\bw) = \Xi(a,\ba) + {{i\tau'}\over {2\tau_2}}(w-\bw) A_w(a,\ba).
$$
We will set $\Xi(a,\ba)=0$.

Because $K_3$ is Hyper-K\"ahler, the curvature
piece in (\ref{curdep}) does not contribute to the instanton
equations.
The holomorphic part of the instanton equations becomes:
\bear
\lam I &=& \px{a}A_w -\px{w}A_a -i A_a\star A_w + i A_w\star A_a
\nn\\ &=&
\px{a}A_w - {{i\tau'}\over {2\tau_2}}A_w
  -2\theta^{aw} (\nabla_a A_w) (\nabla_w A_a)
  +2\theta^{aw} (\nabla_w A_w) (\nabla_a A_a)
\nn\\ &&
  -2\theta^{\ba\bw} (\nabla_\ba A_w) (\nabla_\bw A_a)
  +2\theta^{\ba\bw} (\nabla_\bw A_w) (\nabla_\ba A_a)
\label{instaw}
\eear
We define:
$$
\phi(a,\ba)\equiv -{{i\tau_2 A_w}\over \pi}.
$$
Before we continue with the instanton equation,
let us calculate the periodicity condition because of gauge
invariance.
We pick a gauge transformation:
$$
\Lambda = e^{ 2\pi i m {{w-\bw}\over {\tau-\btau}}
  -2 \pi i n {{\tau \bw -\btau w}\over {\tau-\btau}}}
$$
and calculate the change under:
$$
A_w\rightarrow
 i\Lambda^{-1}\star\px{w}\Lambda +\Lambda^{-1}\star A_w\star \Lambda.
$$
We find:
\bear
\phi &\rightarrow& 
\phi + m+n\btau
+{{\pi\bvth}\over {2\tau_2^2}} (m+n\tau)^2\btau'
\nn\\ &&
 -{{2\pi i\vth}\over {\tau_2}}(m+n\btau)\px{a}\phi
 +{{2\pi i\bvth}\over {\tau_2}} (m+n\tau)\px{\ba}\phi
 +{{\pi\bvth}\over {\tau_2^2}} (m+n\tau)\btau' \phi
 + O(\vth)^2.
\nn
\eear
These equations can be interpreted as follows.
The periodicity equation for $\phi$ can be summarized
in the relation:
\be\label{idua}
(u,a+\delta a)\sim (u+m + n\tau+\delta u,a),
\ee
where:
\bear
\delta a &\equiv& 
 -{{2\pi i\vth}\over {\tau_2}}(m+n\btau) + O(\vth)^2,
\nn\\
\delta u &\equiv&
  {{\pi\vth}\over {2\tau_2^2}} (m+n\btau)^2\tau'
 +{{\pi\vth}\over {\tau_2^2}} (m+n\btau)\tau' u + O(\vth)^2,
\nn
\eear
This is chosen so that a curve given by the equation
$\bu=\phi(a,\ba)$ will be mapped by $\sim$ to the curve
given by:
\bear
\bu &=&
\phi + m+n\btau 
+{{\pi\bvth}\over {2\tau_2^2}} (m+n\tau)^2\btau'
\nn\\ &&
 -{{2\pi i\vth}\over {\tau_2}}(m+n\btau)\px{a}\phi
 +{{2\pi i\bvth}\over {\tau_2}} (m+n\tau)\px{\ba}\phi
 +{{\pi\bvth}\over {\tau_2^2}} (m+n\tau)\btau' \phi
 + O(\vth)^2.
\nn
\eear
Now we can change coordinates to $(a_0,u_0)$ with:
\bear
u_0 &\equiv&
 u +{{\pi\vth}\over {2\tau_2^2}} (\bu^2-2u\bu)\tau' + O(\vth)^2,
\nn\\
a_0 &\equiv&
 a -{{2\pi i\vth}\over {\tau_2}}\bu + O(\vth)^2,
\nn
\eear
In terms of the new coordinates $(a_0,u_0)$, the identifications
(\ref{idua}) become:
$$
(a_0,u_0) \sim (a_0,u_0+m+n\tau(a_0)).
$$
Thus, the identification (\ref{idua}) defines, locally, a manifold
that seems locally identical to the original $K_3$, but with
a different metric!

It has an integral homology class, $F'$,  generated by the new fiber
$$
a_0 =
 a -{{2\pi i\vth}\over {\tau_2}}\bu + O(\vth)^2 = {\mbox{const}}.
$$
The base , $B$, is given by the equation $u_0=0$ (which is the
same as $u=0$).

Now we can proceed with (\ref{instaw}).
Expanding to $O(\theta)$, we find that (\ref{instaw}) becomes:
\bear
\lambda &=& \px{a}A_w - {{i\tau'}\over {2\tau_2}}A_w
-{{i\bvth\tau'}\over {\tau_2}}A_w\px{\ba}A_w.
\nn
\eear
In terms of $\phi$ this reads:
\bear
0 &=&
\px{a}\phi
+{{i\lambda\tau_2}\over {\pi}}
-{{\pi i\bvth |\tau'|^2}\over {2\tau_2^3}}\phi^2
+{{\pi\bvth\tau'}\over {\tau_2^2}}\phi \px{\ba}\phi+O(\vth)^2.
\nn
\eear
Now consider the curve given by the equation:
$$
\Xi(u,a,\ba)\equiv\bu-\phi(a,\ba) = 0.
$$
Define the differential operators:
\bear
D_1 &\equiv&
 \px{a}
 -{{i\lambda\tau_2}\over {\pi}}\px{\bu}
 +{{\pi i\bvth |\tau'|^2}\over {2\tau_2^3}}\bu^2\px{\bu}
 +{{\pi\bvth\tau'}\over {\tau_2^2}}\bu \px{\ba},
\nn\\
D_2 &\equiv&
  \px{u}.
\nn
\eear

The instanton equation implies that, restricted
to the curve $\Xi=0$,
$$
D_1\Xi|_{\Xi=0} = D_2\Xi = 0.
$$
We can rewrite $D_1$ and $D_2$ in terms of $u_0$ and $a_0$ as:
\bear
D_1 &=&
-\left(1+{{\pi \vth\tau'}\over {\tau_2^2}}\bu_0 \right)\px{a_0}
+{{\pi \bvth\tau'}\over {\tau_2^2}}(u_0-\bu_0) \px{\bz_0}
-\left(
{{\pi i\vth}\over {2\tau_2^3}} {\tau'}^2
+{{\pi\vth}\over {2\tau_2^2}} \tau''
\right) ({\bu_0}^2-2u_0\bu_0)\px{u_0}
\nn\\ &&
-{{\pi i\bvth |\tau'|^2}\over {2\tau_2^3}} (u_0-\bu_0)^2 \px{\bu_0}
+{{i\lambda\tau_2}\over {\pi}}\px{\bu_0}
+ O(\vth)^2
\nn\\
D_2 &=&
+{{2\pi i\bvth}\over {\tau_2}} \px{\ba_0}
+\left(
1 -{{\pi\vth\tau'}\over {\tau_2^2}} \bu_0
\right)\px{u_0}
-{{\pi\bvth\btau'}\over {\tau_2^2}} (u_0-\bu_0)\px{\bu_0}.
\nn
\eear
We can now find a modified complex structure on the $K_3$ such 
that $\Xi$ will be a holomorphic curve.
This means that
in terms of the modified complex structure, $J_k^l +\delta J_k^l$,
the equations $D_1\Xi = D_2\Xi=0$ 
should imply $(\delta_k^l- i J_k^l-i \delta J_k^l)\px{l}\Xi = 0$.
In other words,
$\px{u_0} - {i\over 2}\delta J_{u_0}^l\px{l}$ and 
$\px{z_0} - {i\over 2}\delta J_{z_0}^l\px{l}$ should be local
linear combinations of $D_1$ and $D_2$.
We can therefore calculate:
\bear
\delta J_{a_0}^{\ba_0} &=& {{2\pi i\bvth\tau'}\over {\tau_2^2}}(u_0-\bu_0),
\nn\\
\delta J_{a_0}^{\bu_0} &=& 
  -{{\pi\bvth |\tau'|^2}\over {\tau_2^3}}(u_0-\bu_0)^2
  +2\lam\tau_2,
\nn\\
\delta J_{u_0}^{\ba_0} &=& -{{4\pi\bvth}\over {\tau_2}},
\nn\\
\delta J_{u_0}^{\bu_0} &=& 
   -{{2\pi i\bvth\btau'}\over {\tau_2^2}}(u_0-\bu_0).
\nn\\ &&
\label{delJ}
\eear
It is easy to check that $\delta J_k^l$ is invariant under
$u_0\rightarrow u_0 + m + n\tau(a_0)$.

We can also calculate the modified  covariantly constant $(2,0)$ form.
Writing it as $\omega + \delta\omega$, we find:
\bear
\delta\omega_{a_0\ba_0} &=&
  -{i\over 2}\delta J_{\ba_0}^{u_0}
  ={{i\pi\bvth |\tau'|^2}\over {2\tau_2^3}}(u_0-\bu_0)^2   -i\blam\tau_2,
\nn\\
\delta\omega_{u_0\bu_0} &=&
  {i\over 2}\delta J_{\bu_0}^{a_0}  =-{{2\pi i\vth}\over {\tau_2}},
\nn\\
\delta\omega_{u_0\ba_0} &=&
   {i\over 2}\delta J_{\ba_0}^{a_0}
  = -{{\pi\vth\btau'}\over {\tau_2^2}}(u_0-\bu_0),
\nn\\
\delta\omega_{a_0\bu_0} &=&
  -{i\over 2}\delta J_{\bu_0}^{u_0}
 = -{{\pi\vth\tau'}\over {\tau_2^2}}(u_0-\bu_0).
\nn
\eear

We can calculate:
\be
\int_{F'}\delta\omega = -2\pi i\vth,\qquad
\int_{B'}\delta\omega = -i\blam.
\label{FpBp}
\ee

We can now find the requirement on
$\lam$ in (\ref{instaw}) such that a curve
in the homology class:
$$
\hSigma = k\lbrack F'\rbrack + q\hB',
$$
will be analytic in the complex structure given by (\ref{delJ}):
$$
0=\int_\hSigma\delta\omega = -2\pi i k\vth -i q\blam.
$$
This implies:
$$
\lam = -{{2\pi k}\over {q}}\bvth + O(\theta)^2.
$$




\section{Relation with little-string theories}\label{lst}
Instantons on a noncommutative $\MT{4}$ are the solution
to  the Coulomb branch moduli 
space of certain 2+1D theories with $\SUSY{4}$ supersymmetry \cite{CGKM}.

In the following sections we will make an analogous statement
about instantons on a noncommutative $K_3$.
We will argue that the moduli space of instantons on a noncommutative
$K_3$ provides
the low-energy description of a certain
2+1D theory with $\SUSY{4}$ supersymmetry.
In general, the low-energy description of such theories
is a supersymmetric $\sigma$-model on
a hyper-K\"ahler manifold. The dimension of the manifold is
the number of low-energy bosonic fields in the theory.

The theories that we consider are 5+1D theories
compactified on $\MT{3}$. These 5+1D theories
are the two heterotic little-string theories (LST) defined
in \cite{SeiVBR} as the decoupled theories on NS5-branes
of the heterotic string theory in the limit of zero coupling
constant.

Let us take $M_s$ to be the string-scale (the scale of the LST)
and take $R_1,R_2,R_3$ to be the radii of the $\MT{3}$.
For simplicity, we assume that the $\MT{3}$ is of the form
$\MS{1}\times\MS{1}\times\MS{1}$.

The heterotic LSTs have $\SUSY{(1,0)}$ supersymmetry. They also
have an $E_8\times E_8$ or $Spin(32)/\BZ_2$ global symmetry
inherited from the string-theory.
In addition, they possess a global $Spin(4)=SU(2)_L\times SU(2)_R$
symmetry corresponding to rotations in directions transverse
to the NS5-branes. The $SU(2)_R$ does not commute with the
supersymmetry generators and is therefore an R-symmetry.
The $SU(2)_L$ subgroup commutes with the SUSY generators and
is an additional global symmetry.

When we compactify on $\MT{3}$ we have to specify the Wilson
lines corresponding to the global symmetries.
Since we wish to preserve supersymmetry we will not put
any Wilson lines for the R-symmetry.
We therefore have to specify $3\times (16+1)$ Wilson line
parameters.\footnote{We will not discuss compactifications
without vector structure as in \cite{LMST,Bian,WitV}.}
We will refer to the 3 global $SU(2)_L$ Wilson line
parameters  as the {\bf $\alpha$-twists}.

\subsection{Gauge theory limits}
There are various limits of the parameters $R_i$ for which the
problem reduces to an ordinary gauge-theory question.
We can obtain a guage theory limit by using one of two facts:
\begin{itemize}
\item
The low-energy description of the 5+1D $Spin(32)/\BZ_2$ LST
is given by an $Sp(k)$ gauge theory with a global $Spin(32)/\BZ_2$
symmetry and hyper-multiplets in the $(\rep{2k},\rep{32})$ representation
of $Sp(k)\times Spin(32)$
and the anti-symmetric $(\rep{k(2k-1)},\rep{1})$ representation
\cite{WitSMI}.
\item
For $k=1$,
the low-energy limit of the 5+1D $E_8$ CFT compactified on $\MT{2}$
is in general described by a strongly coupled 3+1D CFT \cite{GMS}.
For appropriately chosen $E_8$ Wilson lines
we can get $SU(2)$ QCD with $N_f = 0,\dots,8$ (\cite{GanC,GMS}).
For $k>1$ and appropriately chosen Wilson lines we get
the same $Sp(k)$ theory as above \cite{DLS}.
\end{itemize}

We can use this to generate 3+1D gauge theories as follows.
Compactifying the 5+1D $Spin(32)/\BZ_2$ LST on $\MS{1}$ to 4+1D we
take $M_s\rightarrow\infty$ with $M_s R_1\rightarrow\infty$
and $R_1\rightarrow 0$. We can pick a Wilson-line $W\in Spin(32)/\BZ_2$
such that a field in the fundamental $\rep{32}$ will have
anti-periodic boundary conditions. We can combine it with
a global $SU(2)_L$ Wilson-line (the ``twists'' discussed at
the end of the previous subsection)
such that all hypermultiplets get an extra $(-)$ factor in the boundary
condition. In this way we can get a 4+1D $Sp(k)$ gauge theory
with either some $\rep{2k}$ hypermultiplets
(if we turn on both $W$ and the twist)  or a $\rep{k(2k-1)}$
hypermultiplet (if we just turn $W$ on) or both (if
we turn neither $W$ nor the twist on) or none (if we turn
on only the twist)!
By controlling the value of the Wilson lines we can give small
masses to any of these hypermultiplets.
We can now compactify to 3+1D or 2+1D.



In \cite{CGKM}, the equivalence of the moduli space on the Coulomb
branch of the twisted compactified type-II little-string theory
and the moduli space of noncommutative instantons on $\MT{4}$ was
argued by embedding the NS5-branes inside a Taub-NUT space and
then using a duality transformation to map the system to a system
of D6-branes and a D2-brane.
We will now use a similar technique, adapted to the heterotic 
theory.

\subsection{Embedding in a Taub-NUT space}
A Taub-NUT space is a circle-fibration over $\MR{3}$ such
that the radius of the circle shrinks to zero at the origin.
The metric is:
$$
ds^2 = \rho^2 U(dy - A_i dx^i)^2 + U^{-1} (d\vec{x})^2,\qquad
i=1\dots 3,\qquad 0\le y\le 2\pi.
$$
where $\rho$ is the radius of the circle at $|\vec{x}|=\infty$ and,
$$
U = \left(1 + {{\rho}\over {2|\vec{x}|}}\right)^{-1}.
$$
$A_i$ is the gauge field of a monopole centered at the origin.

The property of the Taub-NUT space that is important for us 
is that the origin $|\vec{x}|=0$ is a smooth point and is a fixed-point
of the $U(1)$ isometry $y\rightarrow y+\epsilon$. This $U(1)$ isometry
acts linearly on the tangent-space at the origin.
In fact, it acts as a $U(1)\subset SU(2)_L$ subgroup of the rotation
group $Spin(4) = SU(2)_L\times SU(2)_R$ that acts on the tangent space
at the origin. The $(-1)\in Spin(4)$ that maps to the identity
in $SO(4)$ corresponds  to $y\rightarrow y+\pi$.

Now, if we place $q$ NS5-branes at the center of the Taub-NUT space and
take $\rho$ to be very large, we can realize an $\a$-twist 
as a shift $y\rightarrow y+\a$.
When we compactify on $\MS{1}$ with coordinate $0\le\theta\le 2\pi$,
we can realize an $e^{i\a}\in U(1)\subset SU(2)_L$ twist,
by identifying:
$$
(\theta, \vec{x}, y) \sim (\theta+2\pi, \vec{x}, y+\a).
$$
This is a Dehn twist of the $y$-circle over the $\theta$-circle.
It can be set as a boundary condition at $|\vec{x}|=\infty$.

Now let us compactify on $\MT{3}$ with 3 $\a$-twists.
In this setting, we take a Taub-NUT space and $q$ NS5-branes.
Let us denote the coordinates as follows:

\begin{tabular}{rcccccccccc}
Object    & 0 & 1 & 2 & 3 & 4 & 5 & 6 & 7 & 8 & 9 \\ \hline
Taub-NUT: & - & - & - & - & - & - & $y$ & $x_1$  & $x_2$  & $x_3$  \\
NS5:      & - & - & - & - & - & - &   &   &   &   \\
$T^3$:    &   &   &   & - & - & - &   &   &   &   \\
\end{tabular}

Here `-' denotes a direction which the object fills and $y$
denotes the Taub-NUT circle direction. The $\a$-twists become
Dehn-twists of the $6^{th}$ circle along the $3^{rd}$, $4^{th}$ and
$5^{th}$ direction, just like \cite{CGKM}.

Now let us consider the sphere $|\vec{x}| = R$ for $R\rightarrow\infty$.
The circle in the $6^{th}$ direction is fibered over it nontrivially
and the $\MT{3}$ is fibered over it trivially.
Let us combine the $6^{th}$ circle to the $\MT{3}$ to form
a $\MT{4}$. This $\MT{4}$ is fibered over the sphere $|\vec{x}| = R$
and we can adiabatically use the S-duality between heterotic string
theory on $\MT{4}$ and type-IIA on $K_3$ to replace the background
with a type-IIA background that at infinity looks like a $K_3$
fibered over the sphere $|\vec{x}| = R$.
The Taub-NUT and NS5-branes in the original heterotic theory
become other BPS objects at the center of the space in the type-IIA
theory. By analyzing what the charge that corresponds to the Taub-NUT
and NS5-branes transforms into
under S-duality, we can determine what these
objects are. Moreover, the $\a$-twists map to some fluxes in the 
type-IIA theory. Our goal now is to determine what these fluxes are
and what are the objects that the Taub-NUT and NS5-branes turn into.

Recall, that if one starts from $k$ NS5-branes inside a $q$-centered
Taub-NUT, then the total NS5-flux is equal to $k-q$.
To see this one should recall the Bianchi Identity for 3-form
field-strength $H$:
$$
{1\over {2\pi}} dH =
-\frac{1}{8\pi^2} trR\wedge R +k\delta(\vec x) \delta(y),
$$
where we did not write the contribution from gauge fields
since they are  present in our consideration only as Wilson lines.
 Integrating this equation over $S^3$ at infinity one gets
 ${1\over {2\pi}}\int_{S^3}{H}=k-q$.

\subsection{Using the S-duality: $IIA/K_3\Leftrightarrow H/\MT{4}$}
\label{revsd}
Let us consider the $\MT{4}$ (in directions $3,4,5,6$) that 
is fibered over the sphere $|\vec{x}|=R$ at $R\rightarrow\infty$.
Let us take 3 $\alpha$-twists denoted by $\alpha_m$ ($m=1\dots 3$).
The metric on $\MT{4}$, written in block form with the blocks of
size $1$ and $3$, is:
\begin{equation}
G_{kl}=\left ( \begin{array}{cc}
R^2 & -\alpha_m R^2\\
-\alpha_n R^2 & g_{nm} +\alpha_n \alpha_m R^2\\
\end{array} \right ),
\qquad
G^{kl}=\left ( \begin{array}{cc}
\frac{1}{R^2} +\alpha^k \alpha_k & \alpha^p\\
\alpha^m & g^{mp} \\
\end{array} \right )
\label{metric}
\end{equation}
Here $k,l=6,3,4,5$. $g_{mn}$ ($1\le m,n\le 3$) is the metric
on the $\MT{3}$ in directions $3,4,5$. $R$ is the radius of the
$6^{th}$ direction and $\alpha^n=g^{nm} \alpha_m$.
The Lagrangian for vector fields in $R^{2,1} \times R^3$ has the form
$$
{\cal L} =
 -{1\over 4} {{\cal F}^i}_{\mu \nu} (M^{-1})_{ij}{\cal F}^{j\mu \nu}.
$$
Here $i=1\dots 24$ and $M\equiv\Omega\Omega^T$
is the $24\times 24$ symmetric matrix
that specifies the point $\Omega$ in the moduli space:
$SO(4,20)/SO(4)\times SO(20)$.
In terms of the physical parameters,
the matrix $M$ is given in \cite{Sch}:
\begin{equation}
 M_{Het} =
\left( \begin{array}{rrr}
G^{-1} &-G^{-1}C & -G^{-1}A^T \\
-C^TG^{-1} & G+C^TG^{-1}C +A^T A & C^TG^{-1}A^T +A^T \\
-A G^{-1}& A G^{-1} +A & \Gamma+A G^{-1} A^T
\end{array} \right).
\end{equation}
It is written in block form, with the blocks of sizes $4+4+16$.
Here $A_k^a$ ($k=6,3,4,5$ and $a=1\dots 16$) are the $E_8\times E_8$
or $Spin(32)/\BZ_2$ Wilson lines
along the $k^{th}$ cycle of the $\MT{4}$.
$\Gamma$ is the Cartan matrix of $E_8\times E_8$ or $Spin(32)/\BZ_2$ and:
\begin{equation}
C_{kl} \equiv {1\over 2} A_k\cdot A_l+B_{kl}\label{C}
\end{equation}
with $A_{k}\cdot A_{l}\equiv {A_k}^b \Gamma_{bc} {A_l}^c$ and $B_{kl}$
is the NSNS B-field.

We will also need to relate a point in the moduli-space,
$\Omega\in SO(4,20)/SO(4)\times SO(20)$
to the physical parameters of type-IIA on $K_3$.
The $24\times 24$ matrix 
$M_{IIA}=\Omega\Omega^T$, is given in \cite{DLM} as:
\begin{equation}
\left( \begin{array}{ccc}
e^\rho & -{1\over 2}e^\rho (b^I b^J d_{IJ}) & e^\rho b^I \\
-{1\over 2}e^\rho (b^I b^J d_{IJ}) &  
e^{-\rho} + b^I b^J d_{IK} {H^K}_J + {1\over 4}e^\rho (b^I b^J d_{IJ})^2 &
-b^K {H^I}_K - {1\over 2}e^\rho b^I (b^K b^L d_{KL}) \\
e^\rho b^J & -b^K {H^J}_K -{1\over 2} e^\rho b^J (b^K b^K d_{KL}) &
{H^I}_K d^{JK} + e^\rho b^I b^J \\
\end{array} \right)
\end{equation}
The matrix is written in blocks of sizes $1+1+22$.
$e^{-\rho}$ is defined as
$e^{-\rho}={M_s}^4 V_{K3}$, where $M_s$ is the string scale of
the IIA-theory and $V_{K3}$ is the volume of K3. 
The vector $b^I$ ($I=1\dots 22$)  specifies the integral
of the NSNS 2-form over a basis of the 22-dimensional
homology $H_2(K_3)$. Choosing a dual basis $\{\omega_I\}_{I=1}^{22}$
of $H^2(K_3,\BZ)$, we can write $B^{NS}=\sum b^I\omega_I$.
The basis of $H_2(K_3,\BZ)$
is chosen such that the intersection matrix is
(we show only non-zero entries):
\begin{equation}
d_{IJ}=
\left(\begin{array}{rrr}
& I_3&\\
I_3 &  & \\
& & \Gamma_{bc}
\end{array}
\right ) \label{int}
\end{equation}
The parameters ${H^I}_J$ describe the metric by
specifying the splitting of $H^2(K_3)$ into self-dual and anti-self-dual
parts. A 2-form $\sum_J \lambda^I\omega_I$ is self-dual if
$\sum_J{H^I}_J\lambda^J = \lambda^J$ and anti-self-dual if
$\sum_J{H^I}_J\lambda^J = -\lambda^J$.
The ${H^I}_J$'s are constrained by the relations:
$$
{H^I}_J {H^J}_K = {\delta^I}_K,\qquad
d_{IJ} {H^J}_K = d_{KJ} {H^J}_I,\qquad
{H^J}_I d_{JK} {H^K}_L = d_{IL},$$ so that 
${H^I}_J$ has 57 independent parameters.

The matrix, $M$,  satisfies:
$$
M^T = M,\qquad M L M^T = L^{-1}
$$
where
$$
L = \left(\begin{array}{cc}
-\sigma^1 & 0 \\ 0 & d_{IJ} \\
\end{array}\right) \quad I=1, \ldots 22.
$$

An S-duality map is given by a $24\times 24$ matrix, $S$,
such that the point in moduli space in the type-IIA theory is
given by:
$$
M_{IIA} = S M_{Het} S^T.
$$
In principle, $S$, is known up to an $SO(4,20,\BZ)$ 
T-duality transformation.

\subsection{The Taub-NUT and NS5-brane}
We chose the matrix $S$ that describes the duality transformation
$M_{IIA}=SM_{Het}S^T$  so as to map the $k-q$ units of NS5-charge
to $k-q$ $\widetilde{D6}$-branes  and map $q$ units of TN-flux
to $q$ D2-branes.
 We use the symbol $\widetilde{D6}$ to denote an object carrying
one unit of $\int_{K_3} F\wedge F$ and  having only pure D6-brane charge
as measured at infinity (and no $H=dB$ flux).
$\widetilde{D6}$-branes can be decomposed
into a D6-brane and a D2-brane,
where now D6 has zero $\int F\wedge F$ flux. D6-branes
are wrapped over $K_3$ and D2-branes are points on $K_3$. 
We will consider the limit of shrinking the $K_3$ to zero
volume  and perform T-duality
on the $K_3$
that interchanges D2 and $\widetilde{D6}$-branes.\footnote{We
are grateful to S. Sethi for useful discussions on this point.}
After T-duality the moduli space is that of 
$k$ instantons in $U(q)$, since:
$$
(k-q)D2+q \widetilde{D6} = kD2 +qD6.
$$

\subsection{The $\a$-twists and instantons on a noncommutative $K_3$}
\label{atwins}
The 80 parameters that specify the $K_3$ and the $B^{NS}$ fluxes
on the type-IIA side are functions of the parameters of the
$\MT{4}$ compactification on the heterotic side.
We now wish to know how the $K_3$ parameters depend on the 
relevant parameters on the heterotic side, 
namely, the $\a$-twists and the metric on $\MT{3}$ (in directions 
$3,4,5$).

We find that the moduli of the metric on $K_3$ 
(i.e. ${H_I}^J$ and $e^{-\rho}$) are independent of the $\a$-twists.
Specifically:
\be\label{Hmn}
\begin{array}{lll}
{H^{m}}_{n}=-g^{np}C_{pm}, &
  {H^{m+3}}_{n}=g^{nm},  &
    {H^d}_{n}=-{A_l}^{d-6} g^{ln}, \\
{H^{m}}_{n+3}=g_{nm}+A_n \cdot A_m +C_{pn}g^{pk}C_{km}, &
  {H^{m+3}}_{n+3}=-C_{pn}g^{pm}, &
    {H^d}_{n+3}= {A_n}^{d-6}+ {A_l}^{d-6} g^{lp}C_{pn}, \\
{H^m}_{c}=A_{m,c-6}+C_{pm}g^{pk} A_{k,c-6}, &
  {H^{m+3}}_{c}= -A_{p,c-6} g^{pm},  &
    {H^d}_c=\delta_c^d +{A_n}^{d-6} g^{nm}A_{m,c-6}, \\
m=1\dots 3, &
m+3=4\dots 6, &
d =7\dots 24. \\
\end{array}
\ee
Here we used the notation of subsection (\ref{revsd}).
Also:
\be\label{rho}
e^{-\rho}=\frac{1}{{m_s}^2 R^2}.
\ee

We can also check that the matrix $S$
transforms the periodic parameters $\alpha_m$ (the twists),
on the Heterotic side into the periodic
moduli $b^m$   ($B^{NS}$ fluxes) on IIA side. 
In the basis of $H_2(K_3,\BZ)$ that corresponds to (\ref{int}), 
we have
\begin{equation}
b^{m}=\alpha_m\, (m=1,2,3), \qquad b^{m+3}=0\,(m+3=4,5,6),
\qquad b^c=0\, (c=7\dots 22).
 \label{b}
\end{equation}
We will now argue that switching on
$\alpha_m$ results in the instantons being noncommutative.

To make this point, let us also recall that in \cite{CDS},
noncommutative  $U(N)$ SYM on $\MT{4}$ was defined  by taking
$N$ D0-branes in $\MT{4}$ with $B^{NS}$ flux and
shrinking the area of $\MT{4} \rightarrow 0$. 
 Analogously, we can take a $K_3$ with $B^{NS}$ flux and 
$q$ D0-branes and we believe that in the limit that
the $K_3$ shrinks to zero size the system is described by
$U(q)$ NCG on a noncommutative $K_3$.

 After S-duality we have  the system
of $q$ D2-branes and $k-q$ $\widetilde{D6}$-branes.
In order to study the moduli space of the system,
we will discuss a system of $q$ D0-branes and $k-q$
$\widetilde{D4}$-branes on $K_3$, where $\widetilde{D4}$-branes
are defined to have only pure D4-charge.

Let us define ${\tilde b}^i$ (for $i=1,2,3$) to be the
components of $B^{NS}$ along
self-dual 2-forms in the decomposition into selfdual and
anti-selfdual parts.
We can show that a  non-zero ${\tilde b}^i$ flux
results in  a bound state with mass:
\begin{equation}
{m^2}_{D0+\widetilde{D4}} < (m_{D0}+m_{\widetilde{D4}})^2, \label{m}
\end{equation}
Here 
${m^2}_{D0+\widetilde{D4}}$ denotes the square of the mass 
of the bound state of $q$ D2-branes  and $N$
$\widetilde{D4}$-branes, and $m_{\tilde D4}$
($m_{D0}$) is the mass of the $N$ $\widetilde{D4}$
($q$ D0) isolated branes.
 
Equation (\ref{m})
means that one cannot separate the D0 and $\widetilde{D4}$-branes in the
presence of a non-zero ${\tilde b}^i$. This in turn implies
that instantons cannot shrink to zero size,
which is a sign of noncommutativity.

We will prove (\ref{m}) for arbitrary numbers $q$ of D0-branes
and $N$ of $\widetilde{D4}$-branes. 
 We start from the fact that $N$ $\widetilde{D4}$-branes
and $q$ D0-branes are characterized
by a Mukai vector $v=(N,0,N-q)$,
where the intersection product was defined
as \cite{RD}:
\begin{equation}
v \cdot v'=\int \left( v^2 \wedge v'^2 -v^0 \wedge v'^4-v'^0 \wedge v^4
\right ), \hspace{1cm} v=\left ( v^0,v^2,v^4 \right ), v^k \in
H^k(K_3).
\label{Mukai}
\end{equation}

As it was done in \cite{MK} and \cite{RD},
we defined vectors that span positive
4-plane in $H^{*}(K_3)$

\begin{equation}
E_0=\left (1,B,
  {1\over 2} B \wedge B 
  -{1\over 2} \sum_i \omega_i \wedge \omega_i \right),\qquad
E_i=\left ( 0, \omega_i, B \wedge \omega_i \right )\, (i=1,2,3).
\label{basis}
\end{equation}
Here $\omega_i$ ($i=1,2,3$) is a basis for the self-dual subspace
of the 22-dimensional $H^2(K_3)$.
We can find the projection of $v$ onto this plane, $v_{proj}$, and
determine the squared mass by $m^2={v^2}_{proj}.$ 
 Expanding the self-dual 2-forms as
$\omega_i=\sum_{J=1}^{22}\epsilon_{(i)}^J\omega_J$, we find:
\begin{equation}
\int \omega_i \wedge \omega_j = d_{IJ}\epsilon_i^I \epsilon_j^J
\label{inters}
\end{equation}
where
\begin{equation}
 \epsilon_{(i)}^J=\left ( \begin{array} {c}
-(C_{kn} +g_{kn}) \delta_{(i)}^k \\
\delta_{(i)}^n \\
-{A_k}^b\delta_{(i)}^k
\end{array} \right )
\label {eps3*}
\end{equation}
and the relation between heterotic  and IIA moduli was given
in (\ref{Hmn}).  
 From (\ref{basis}) and (\ref{eps3*}) it follows that
\begin{equation}
E_0 \cdot E_0 =\lambda,\qquad
E_0 \cdot E_j=0,\qquad
E_i \cdot E_j =2g_{ij}.
\label{basis1}
\end{equation}
with 
$\lambda =2 \tr{g}$.
Finally, we obtain:
\begin{equation}
\displaystyle 
{m^2}_{D0+\tilde D4}=\frac{1}{\lambda} \left
 ({N\over 2} \left ( 2-\lambda +2{\tilde b} g 
{\tilde b} \right ) - q \right )^2 +2N^2 {\tilde b} g {\tilde b},
\label{mass}
\end{equation}
Here  ${\tilde b} g {\tilde b}\equiv {\tilde b}_i g^{ij} {\tilde b}_j$
for $i,j=1,2,3$. It is straightforward to show that 
${m^2}_{D0+\tilde D4} < (m_{D0}+m_{\tilde D4})^2$ for any non-zero
${\tilde b}_i$ and any numbers $N$, $q$ of D4 and D0-branes.

\subsection{Generalized twists in the Heterotic LST}
As pointed out in \cite{SeiVBR},
the little-string theories exhibit T-duality.
Compactification of the heterotic LSTs on $\MT{d}$ is specified
by an external parameter space of $SO(d,16+d)/SO(d)\times SO(16+d)$
(the metric, $B$-field, and Wilson lines) 
and there are discrete T-duality identifications
given by $SO(d,16+d,\BZ)$ which act on the spectrum by exchanging
momenta and winding quantum numbers (and can also mix it with
global $U(1)^{16}$ quantum numbers which is the generic unbroken 
part of the gauge group).
One can ask what happens if we compactify with an $\a$-twist.
In \cite{CGKM} it was argued that the T-dual of an 
$\a$-twist is another kind of twist. For the T-dual ``$\eta$-twist'',
a state with a nonzero $SU(2)_L$ charge also has
a fractional winding number. We would now like to define generalized
twists for the heterotic LSTs, in a similar manner.

In (\ref{Hmn}-\ref{rho}), 
we saw  that the moduli of the metric on $K_3$ are independent of
the values of the $\a$-twists on the heterotic side.
This fact  suggests the following definition of a T-dual of the
$\alpha$ twist. Start from type-IIA using the moduli for the metric on 
$K_3$,  given by equations
(\ref{Hmn}-\ref{rho}),
and take all components of $B^{NS}$ (in the integer basis
of $H^2(K_3,\BZ)$)
to be non-zero $b^I=(\alpha_m, \eta^m, \gamma^c)$.
 Then, apply the transformation, 
given by $S^{-1}$, to the Heterotic theory and find the appropriate
moduli.

For example, we can see that, with this definition of the
generic twists,  the $SU(2)_L$ Wilson lines along the $\MT{3}$
(on which the heterotic LST is compactified) are given, in terms
of $\a_m,\eta^m,\gamma^c$, by:
\begin{equation}
 G^{1m}=\alpha_k g^{mk}-\eta^n g^{mk}C_{kn} -\gamma^d g^{mk} a_{kd}+
{1\over 2}e^{\rho} \eta^m (b^I d_{IJ} b^J),\qquad
G^{nm}=g^{nm}+e^{\rho}\eta^n \eta^m, \label{Utw}
\end{equation}
The other parameters are rather complicated expressions
and we present them in Appendix (\ref{appgtw}).

 The point in moduli-space, on the type-IIA side, depends only on
the self-dual part of the $B$-field, which is specified by
3 parameters. In subsection (\ref{atwins}), we introduced the
notation ${\tilde b}^i$ (for $i=1,2,3$) for the components
of the $B$-field along some fixed basis of the self-dual 2-forms
on $K_3$.
 We are mostly interested in ${\tilde b}^i$ for $i=1,2,3$,
since they affect the moduli space of instantons.
 For the situation with generalized twists, they are given by:
\begin{equation}
{\tilde b}^i={1\over 2} g^{i\, n} \left ( 
-\alpha_n +(C_{nk}+g_{nk}) \eta^k + A_n \cdot \gamma\right)
\label{instantons}
\end{equation}

\subsection{Conclusion}
We conjecture that the moduli space of the heterotic little-string
theory of $k$ NS5-branes
compactified on $\MT{3}$ with $\a$-twists is given by the moduli
space $\MSP_{k,1}$ of $k$ $U(1)$ instantons on a noncommutative $K_3$.
The noncommutativity, $\theta$, is specified by a symplectic form
whose inverse, $\theta^{-1}$ is a closed 2-form on the $K_3$.
Its expansion in terms of an integral basis of $H^2(K_3,\BZ)$ is
given in terms of the 3 $\a$-twists in (\ref{b}).

\section{Generalization to NS5-branes
  at $A_{q-1}$ singularities}\label{gen}
In previous sections we found moduli spaces of instantons
of a noncommutative $U(1)$ gauge theory.
How can we get instantons for $U(q)$ gauge theories with $q>1$?
On the face of it, all we need to do is replace the Taub-NUT
space with a multi-centered Taub-NUT space.
In particular, the classical multi-centered Taub-NUT space
can have, for a particular choice of parameters, an $A_{q-1}$
singularity and we can naively conclude that the moduli space
of the theories that we get by placing heterotic NS5-branes
at $A_{q-1}$ singularities and compactifying on $\MT{3}$ 
is the moduli space of $U(q)$ instantons on $K_3$ as before.

However, unlike the type-II string theories, the moduli spaces
of $A_{q-1}$ singularities and NS5-branes in $A_{q-1}$ singularities,
receive quantum corrections \cite{SenKK,WitKK,Rozali,Krogh}.
For example, the hypermultiplet
moduli space of an $A_1$ singularity, corresponding to
the normalizable metric and 2-form deformations, in the
heterotic string theory is a blow-up of $\MR{4}/\BZ_2$
\cite{SenKK,WitKK} and does not have a singular point that
can be associated with a singular space, even though classically
the moduli space is exactly $\MR{4}/\BZ_2$ and the origin of it
is singular and can be associated with a singular space.

We would like to generalize the question we asked in (\ref{lst})
to a question like: ``what is the Coulomb branch moduli space
of the 5+1D theory of $k$ NS5-branes  at an $A_{q-1}$ singularity
compactified on $\MT{3}$ with Wilson lines.''
In the type-II LST case, the answer was proposed in
\cite{CGKM} to be the moduli space of $k$ 
noncommutative $U(q)$ instantons on $\MT{4}$.

In the heterotic case, the statement of the problem is somewhat
ambiguous since we need to specify at what point in the hyper-multiplet
moduli space we are.
Let us denote this hyper-multiplet moduli space
by $\ModHyp$.
 If there is no singularity in $\ModHyp$,
 then there is no implied ``special'' point as in the
type-II case.

We can replace the $A_{q-1}$ singularity with a Taub-NUT space.
Then, the Taub-NUT space has a $U(1)$ isometry which, as before,
becomes a global symmetry of a decoupled theory on the NS5-branes.
This $U(1)$ also acts nontrivially on the hyper-multiplet moduli
space $\ModHyp$.
If we put nonzero Wilson lines of this $U(1)$ along the $\MT{3}$
(the ``twists'') then we are forced to be at a fixed-point
of this $U(1)$ in $\ModHyp$.
This restricts the choice of points in $\ModHyp$, but there
are still several cases.
We will argue below that these match nicely with properties
of the moduli space of instantons on $K_3$.

\subsection{Review of heterotic NS5-branes at singularities}
\label{amres}
The type of 5+1D low-energy that one gets for $k$ NS5-branes
and an $A_{q-1}$ singularity of the heterotic string theories,
was analyzed in \cite{AspMor}.
They found that for $k< 4$ the low-energy description is the
same as for $q=1$ for both $E_8\times E_8$ as well as $Spin(32)/\BZ_2$
gauge groups.
They also characterized the theories in terms of the local
gauge group $G$ and the number of tensor multiplets, $n_T$.
They found that $k$ $E_8$ NS5-branes always give $n_T\ge k$ 
and $k$ $Spin(32)/\BZ_2$ NS5-branes always give 
$Sp(k)$ as a factor of $G$. For $q\ge 2$ and
$k=4$ $E_8$ NS5-branes one always
has $G=SU(2)$ and $n_T = 4$. For $q\ge 2$ and
$k=4$ $Spin(32)/\BZ_2$ one has $G=Sp(4)$ and $n_T=1$.

\subsection{The hypermultiplet moduli space}
The hyper-multiplet moduli space, $\ModHyp$,
 of $k$ NS5-branes at a $q$-centered Taub-NUT space
comprises, classically, of the positions of the NS5-branes
($4k$ variables), the blow-up modes of the metric on the
Taub-NUT space ($3(q-1)$ parameters) and the NSNS 2-form
modes ($(q-1)$ parameters). As explained in \cite{SenKK,WitKK},
it receives quantum corrections that depend on $\a'$ but not
on the string coupling constant $\lam$.
If we also include the absolute position of the Taub-NUT space
we get $4(k+q)$ dimensions.
It was argued in \cite{Rozali} that $\ModHyp$  is the
same as the Coulomb branch moduli space of 2+1D QCD with
gauge group $SU(q)\times U(1)^k$ and $N_f=k$ massless quarks
that are in the fundamental representation of $U(q)$ and are
charged under one of the $U(1)$'s.
The coupling constant of the $U(1)$'s was argued to be infinite.
In fact, the duality that we are using is the same as that
used by \cite{Rozali} to solve the hyper-multiplet moduli space.

This space has an overall $U(1)$ isometry. 
Physically, 
this $U(1)$ isometry is the isometry of the Taub-NUT space
and acts on the positions of the NS5-branes.
In the 2+1D QCD language, let us consider the diagonal $U(1)$ subgroup
 of $U(1)^k$.
The 2+1D dual of the photon that corresponds to this $U(1)$
can be shifted without changing the metric on the Coulomb branch.
This is the $U(1)$ isometry on the quantum space.

As we have argued, once we compactify on $\MT{3}$ and introduce
generic twists, the hypermultiplet moduli space reduces to the
fixed point locus of the $U(1)$ isometry.
Let us denote this locus by $\ModHypZ$.
 From the classical limit,
We expect it to be of dimension $4q$.
If we ignore the overall center of mass position of the Taub-NUT
space, we get $4(q-1)$ parameters that correspond to the deformation
modes of the Taub-NUT solution.
For example, with one NS5-brane ($k=1$) and $q=2$, 
the Taub-NUT space has two centers where the fibered circle
shrinks to zero. The NS5-brane can be placed in either of these
centers.
In terms of instantons, this would correspond to breaking
$U(2)\rightarrow U(1)\times U(1)$ and placing the instanton
in one of the $U(1)$ factors.
If the instantons were commutative, this configuration
would be supersymmetric. However, for noncommutative instantons
this configuration has to break supersymmetry.
One way to see this is in terms of curves in $K_3$.
The classes $\hB+\hF$ and $\hB$ cannot be simultaneously
analytic unless the $K_3$ is elliptically fibered with a section.
In terms of the twisted generalized LST, this means that
after compactification on $\MT{3}$ with generic twists,
we expect that there is no supersymmetric vacuum.

With $k=1$ and $q=2$, the singularity in the hyper-multiplet moduli space
is smoothed out by quantum corrections and that means that 
we cannot realize
instantons with $k=1$ inside $U(2)$ except by breaking
$U(2)\rightarrow U(1)\times U(1)$.
In terms of holomorphic curves it means that we cannot find
a holomorphic curve of class $2\hB+\hF$ inside $K_3$.

In order to realize instantons inside $U(2)$, we have to have
a special point in $\ModHypZ$ where intuitively, the two centers
of the Taub-NUT coincide. This means that the moduli space 
should have a singularity. Since the moduli space of 
$SU(2)$ QCD with $N_f=k$ is singular for $k\ge 2$, we see that
we expect to have curves of classes $2\hB+k\hF$ for $k\ge 2$,
inside the $K_3$.
This is also the condition that the self-intersection of the
curve should be greated or equal to $(-2)$ which is required for
irreducible curves. This also agrees \cite{VafID}
with the results obtained from the duality between
type-IIA on $K_3$ and heterotic string theory on $\MT{4}$.

\subsection{The Coulomb branch}
After we have placed $k$ NS5-branes at the $q$-centered Taub-NUT space
and taken the decoupling limit $\lambda\rightarrow 0$
with $M_s$ kept fixed, we obtain a 5+1D little-string theory
with a low-energy description that, in general could contain
tensor multiplets, vector multiplets and hyper-multiplets.
We are interested in the Coulomb branch of the theory after
compactification on $\MT{3}$ with twists of the global
$U(1)$ discussed above.
Following the same arguments as in section (\ref{lst}), 
we can conclude that the moduli space is the same as the
moduli space of $k$ instantons of noncommutative $U(q)$ Yang-Mills
theory on $K_3$. The three parameters that specify the
anti-self-dual part of the 2-form that determines the noncommutativity
of the $K_3$ is determined, as before, by the twists.


\section{Seiberg-Witten curves, spectral-curves
 and T-duality}\label{swrel}
As an application of these results we can motivate the spectral-curve
construction of instanton moduli spaces on a noncommutative $K_3$.

We have seen that the moduli space $\MSP_{k,n}(\theta)$ of 
$k$ noncommutative  $U(n)$ instantons on $K_3$ with a noncommutativity
parameter $\theta$, given by an anti-self-dual 2-form on the $K_3$
corresponds to the moduli space of the Coulomb branch of 
a certain 5+1D theory compactified on $\MT{3}$.

Let us take a special $\MT{3}$ of the form $\MT{2}\times \MS{1}$
and let us take the limit that the radius $R$, of $\MS{1}$, is very
big. We can then compactify in two steps. The first step is 
to obtain a 3+1D theory by compactifying the 5+1D theory on $\MT{2}$.
The low-energy of this theory is described by a certain
Seiberg-Witten curve of genus $g$. When $R\rightarrow\infty$
we can compactify the low-energy effective 3+1D action on $\MS{1}$.
This procedure was described in \cite{SWThreeD}, and we get a
a hyper-K\"ahler moduli space that is described as the collection
of Jacobian varieties of all the Seiberg-Witten curves.
Recall that all the Seiberg-Witten curves form a $g$-dimensional
(complex) space and the Jacobian $\MT{2g}$, of each curve,
 is also a $g$-dimensional complex space.
Together we get a $2g$-dimensional
space which is hyper-K\"ahler.

In the context of instanton moduli spaces, the Seiberg-Witten curves
are called the ``spectral-curves'' and the points on the Jacobian
are called the ``spectral-bundle.''

As an example, let us describe how these considerations  translate
to the spectral-curve construction of noncommutative instantons
on $\MT{4}$.
According to \cite{CGKM}, the moduli space $\MSP_{k,n}$
of instantons on a noncommutative $\MT{4}$ is equivalent to the
moduli space of
little string theories on $\MT{3}$ with twists.
The $\MT{2}$ can always be written (in several ways) as
a $\MT{2}$-fibration over a base $\MT{2}$.
Let us denote the base by $B$ and let $z$ be a doubly-periodic
coordinate on $B$ such that $z\sim z + e_1$ and $z\sim z+ e_2$.
Let $w_a(z)$ ($a=1\dots n$) be the local map from the base to 
the fiber $\MT{2}$.

If we take the little string theories on $\MT{2}\times \MS{1}$
with $\MS{1}$ very large, we get a $\MT{2}$-fiber that is very small.
The little-string theories moduli space can also be deduced
using the construction in \cite{WitBR} of $n k$ D4-branes
suspended between pairs of cyclic $n$ NS5-branes.
In this limit, the twists correspond to Dehn twists as in the
elliptic models of \cite{WitBR}.

Thus we can conclude that if the noncommutativity is described
by a 2-form
$\theta_{i I}$ with $i=1,2$ a coordinate on the base $\MT{2}$
and $I=1,2$ a coordinate on the fiber $\MT{2}$, then the spectral
curves are $N$-fold maps from the base to the fiber with
twisted boundary conditions given by:
$$
w_a(z+\sum_{j=1}^2 N_j e_j) = w_a(z) + \sum_{I=1}^2\theta_{i I} b_I.
$$
Here $b_1,b_2$ are a basis of the lattice such that the fiber
$\MT{2}$ is given by:
$$
w\sim w+b_1\sum w+b_2.
$$

\subsection{Relation to T-duality}
The relation between moduli spaces of noncommutative instantons
on $\MT{4}$ and $K_3$ and spectral curves also follows from
T-duality of type-IIA on these spaces.
This is a generalization of the commutative case \cite{VafID},
and it was explicitly constructed for $\MT{4}$ in \cite{KKKLLY}.
Let us briefly recall how this works for $K_3$.
We start with the definition of noncommutative $U(q)$ gauge theory
on $K_3$, in the spirit of \cite{CDS,DH}.
Namely, we take $q$ D0-branes and $k$ D4-branes
on  $K_3$ and send the volume of  $K_3$ to zero
while keeping a constant NSNS $B$-field flux.
We now perform a T-duality transformation that transforms
the $K_3$ into another $K_3$ that we can present
as an elliptic fibration (not necessarily
with a section) and such 
that a D0-brane turns into a D2-brane wrapped on the fiber
and a D4-brane turns into a D2-brane wrapped on the base.
The base and fiber correspond to $H_2(K_3,\BZ)$ classes
but they are not necessarily analytic in the same complex structure.
Now the $k+q$ D2-branes form a single D2-brane that is analytic
in some complex structure.

Let us describe the original $K_3$ (with the $q$ D0-branes
and $k$ D4-branes) as in (\ref{basis}).
We write it as:
$$
E_0=(1,0,0,\int_{C_3} B,\dots,\int_{C_{22}} B,
  \int {1\over 2} B \wedge B -V),\qquad
E_J=( 0,S_1,S_2, \int_{C_3} J,\dots,\int_{C_{22}} J, 0)$$ 

$$E_{\omega}=( 0, 0, 0,\int_{C_3}\omega,\dots,\int_{C_{22}} \omega,
B \wedge \omega ), \qquad 
E_{\overline{\omega}}=( 0, 0, 0,\int_{C_3}\overline{\omega},
\dots,\int_{C_{22}} \overline{\omega},
B \wedge \overline{\omega })\,
$$
where V is the volume of $K_3$ and
$C_1,\dots,C_{22}$ is a basis for $H_2(K_3,\BZ),$
which we choose  such that $C_1$ corresponds to the fiber
of the elliptic fibration, and $C_2$ corresponds to the base.
The intersection numbers are:
$$
C_1\cdot C_2 = 1,\qquad
C_1\cdot C_1 = 0,\qquad
C_2\cdot C_2 = -2.
$$
We used the fact that $\int_{C_1}\omega=\int_{C_2}\omega =0$,
( the fiber and the base of the original $K_3$ are
analytic in the chosen complex structure) and that we are
dealing with NSNS flux B having only (2,0) and (0,2) parts. 
We also introduced the notation for the volume of the fiber
and the base respectively
 $S_1\equiv\int_{C_1} J, \qquad S_2\equiv\int_{C_2} J, $
 with J being Kahler form on the $K_3.$
After T-duality the vectors which span a positive-definite
4-plane in $ R^{(4,20)} $ become:
\bear
'E_0 &=& (0,1,
 \int {1\over 2} B \wedge B -V, 
\int_{C_3}B,\dots,\int_{C_{22}} B, 0),\nn\\
'E_{J'}
 &=& (S_1,0,0
 \int_{C_3}J,\dots,\int_{C_{22}} J, S_2),\nn\\
'E_{\omega'} &=& (0,0, \int B\wedge \omega,
 \int_{C_3}\omega,\dots,\int_{C_{22}} \omega, 0)\nn
\eear
After interchanging $'E_0 $ and $'E_{J'}$
we can read off the following information:
\bear
\int_{C_{1}'}J'=1 & \int_{C_{2}'}J'=\half \int B \wedge B
\nn\\
\int_{C_1'}\omega'=0 & \int_{C_2'}\omega'=\int B \wedge \omega,
\nn\\ 
\int_{C_1'}\overline{\omega'}=0 &
\int_{C_2'}\overline{\omega'}=\int B \wedge \overline{\omega }. \nn
 \eear
We want to find  a complex structure on a dual $K_3$
such that  curve $\Sigma=q[C_2']+k[C_1']$ be analytic.
So, we define a new  form
$$
\Omega=aJ'+b\omega' +c \overline{\omega'}
$$
and impose three conditions:
$$
\int_{\Sigma} \Omega=0,\qquad
\int_{K_3'} \Omega \wedge \Omega =0,\qquad
\int_{K_3'} \Omega \wedge \overline{\Omega} =
{\mbox{Vol}}(K_3').
$$ 
The first condition results in
$$
a\left(q\left(\half \int B\wedge B -V\right)+k\right)+
 bq\int B\wedge\omega +c q\int B\wedge \overline{\omega}=0.
$$
At this point we take the limit of $V\rightarrow 0$
as in \cite{CDS,DH,SWNCG}.
We obtain:
\be\label{aqkBB}
a\left(\frac{q}{2}\int B\wedge B +k\right)+
 b q\int B\wedge\omega +c q\int B\wedge \overline{\omega}=0.
\ee
The second and third conditions give:
$$
a^2 +2bc=0,\qquad |a|^2 + |b|^2 + |c|^2=1.
$$

Now let us take the limit that $\theta$ is small and compare
the $O(\theta)$ term found in (\ref{FpBp}) with the present
calculation.
We assume that the $B$-field on (the original) $K_3$  takes
the form:
$$
2\pi i B= \vth^{-1}\omega -\bvth^{-1}\overline{\omega}.
$$
This equation is the analog of $\theta=B^{-1}$ in the 
flat space case of \cite{SWNCG}. In order to be consistent
with the conventions of \cite{SWNCG}
of assigning dimensions to parameters, namely that $B$ and $\theta$
are dimensionless and only the metric is dimensionful, we have to
assume that $\omega$ is dimensionless and is normalized such that
$\int\omega\wdg\overline{\omega}$ is kept fixed.
This implies that $a=O(\vth)$, $b=1+O(\vth)$ and $c=O(\vth)^2$.
To check the relation between the NSNS flux $B$
on the original $K_3$ an
the noncommutativity parameter $\vth$ we compare
$\int_{C_1'} \Omega=a$ with 
$\int_{F'}\delta \omega =-2\pi i \vth$ (obtained in (\ref{FpBp}))
and find agreement.
Note that for small $\vth$ the period $\int_{F'}\delta\omega$
is independent of $k$ and $q$. However, from equation (\ref{aqkBB})
we expect higher order terms in $\vth$ to depend on $k$ and $q$.


\section{Summary and further directions}
Let us summarize our results:
\begin{itemize}
\item
We have explicitly constructed instantons on a noncommutative $\MT{4}$
in terms of spectral curves.
The spectral curves are constructed inside a modified
$\MT{4}$ given by (\ref{newbf}).
\item
We wrote down the instanton equations on a noncommutative manifold,
to first order in the noncommutativity. We have seen that for a generic
metric there is an extra curvature dependent term that vanishes
for hyper-K\"ahler manifolds.
\item
We have explicitly constructed instantons on a noncommutative $K_3$,
to first order in the noncommutativity, in terms of spectral
curves. The spectral curves are constructed inside a modified
$K_3$ with a complex structure given by (\ref{delJ}).
\item
We have argued that the moduli spaces of noncommutative $U(1)$
instantons on $K_3$ are the Coulomb branch moduli spaces of
compactified heterotic little-string theories and the
moduli spaces of noncommutative $U(q)$ instantons on $K_3$
are the Coulomb branch moduli spaces of compactified generalized
heterotic little-string theories obtained from heterotic
NS5-branes at $A_{q-1}$ singularities.
\end{itemize}

Let us suggest some further directions that might be interesting
to study:
\begin{itemize}
\item
Expand the instanton equations on a curved
manifold to all orders in the noncommutativity parameter
or the curvature and prove the spectral curve construction
of instantons on a noncommutative $K_3$ directly (rather
than from the T-duality argument).
\item
Explore the relation between the moduli spaces
of noncommutative instantons on $K_3$ and observations of Aspinwall
and Morrison \cite{AspMor} about the theories of heterotic 
NS5-branes at $A_{q-1}$  singularities.
For example, the observation that
for $k< 4$ the low-energy description is the
same as for $q=1$ is related to the fact that for $k<4$
and $q>1$ the commutative instantons are reducible and
therefore the moduli spaces of the theory compactified
on $\MT{3}$ (without twists) decomposes into a product.
(Recall from (\ref{geomef}) that if $k< 2q$, we cannot find an
irreducible curve.)

\end{itemize}
\section*{Acknowledgments}
We are very grateful to M. Berkooz, K. Intriligator, V. Periwal,
P. Schupp and S. Sethi for discussions.
The research of OJG is supported by NSF grant number PHY-9802498.
The research of AYuM is supported by the Russian Grant for
support of scientific schools No. 00-15-96557 and by the 
RFFI Grant No. 00-02-16477.

\appendix
\section{Appendix}\label{appgtw}
In section 4 we have presented  $G^{mn}$ and $G^{1m}$ parameters,
see eq.(\ref{Utw}), which correspond to U-dual of $\alpha $-twist.
Here we will complete the list of Heterotic moduli in the presence
of the generic twist.
\begin{equation}
 G^{11}=e^{-\rho} +b^I d_{IK} {H^K}_J b^{J} 
+\frac{1}{4} e^{\rho} (b^I d_{IK} b^K)^2,
\end{equation}
where ${H^K}_J$ are taken from (\ref{rho}), $d_{IJ}$ from (\ref{int})
and 
$b^I=(\alpha_m, \eta^m, \gamma^c) .$
\begin{equation}
a'_{kb}={\cal K}^{-1}_{kn}
  \Bigl ( G^{11}({H^n}_b+ e^{\rho}\alpha_n \gamma_b)
-G^{n1}({H^b}_Jb^J+{1\over 2}e^{\rho}\gamma^b b^I d_{IJ} b^J) \Bigr),
\end{equation}
where ${\cal K}^{nk}=G^{n1}G^{1k}-G^{11}G^{nk}.$
Let us draw your attention to the fact that in the presence of the
generic twist Wilson lines over the circles of $T^3$ are different
from what one has without twists. Only for $\alpha$-twist
$a'_{kb}=a_{kb}.$

In the generic case one has to introduce Wilson line over
the Taub-Nut circle at infinity
\begin{equation}
\displaystyle
a_1^b=-\frac{{H^b}_Jb^J+{1\over 2}e^{\rho}\gamma^b b^I d_{IJ} b^J
 +G^{1n}{a'_n}^b}{G^{11}}
 \end{equation}
and switch on non-zero $B_{n1}$ field
\begin{equation}
B_{n1}=(e^{\rho}\eta^{m} -{1\over 2} a_1\cdot a_1 G^{1m}){\cal N}_{mn}
-{1\over 2}a'_n \cdot a_1,
\end{equation}
where ${\cal N}_{mn}=(G^{mn})^{-1}$.

One gets a complicated expression for the $B'_{mn}$ field.
It isalso different from the $B_{mn}$ field, present before the twist:
\begin{equation}
B'_{kn}=-{\cal N}_{km} \Bigl ( {H^n}_m + e^{\rho}\alpha_n \eta^m
+ (B_{1n}+{1\over 2} a_1 \cdot a'_n)G^{1m} \Bigr ) 
  -{1\over 2} a'_k \cdot a'_n
\end{equation}
\def\np#1#2#3{{\it Nucl.\ Phys.} {\bf B#1} (#2) #3}
\def\pl#1#2#3{{\it Phys.\ Lett.} {\bf B#1} (#2) #3}
\def\physrev#1#2#3{{\it Phys.\ Rev.\ Lett.} {\bf #1} (#2) #3}
\def\prd#1#2#3{{\it Phys.\ Rev.} {\bf D#1} (#2) #3}
\def\ap#1#2#3{{\it Ann.\ Phys.} {\bf #1} (#2) #3}
\def\ppt#1#2#3{{\it Phys.\ Rep.} {\bf #1} (#2) #3}
\def\rmp#1#2#3{{\it Rev.\ Mod.\ Phys.} {\bf #1} (#2) #3}
\def\cmp#1#2#3{{\it Comm.\ Math.\ Phys.} {\bf #1} (#2) #3}
\def\mpla#1#2#3{{\it Mod.\ Phys.\ Lett.} {\bf #1} (#2) #3}
\def\jhep#1#2#3{{\it JHEP.} {\bf #1} (#2) #3}
\def\atmp#1#2#3{{\it Adv.\ Theor.\ Math.\ Phys.} {\bf #1} (#2) #3}
\def\jgp#1#2#3{{\it J.\ Geom.\ Phys.} {\bf #1} (#2) #3}
\def\cqg#1#2#3{{\it Class.\ Quant.\ Grav.} {\bf #1} (#2) #3}
\def\hepth#1{{\it hep-th/{#1}}}



\begin{thebibliography}{99}


\bibitem{CDS}{A.~Connes, M.R.~Douglas and A.~Schwarz,
  {``Noncommutative Geometry and Matrix Theory:
  Compactification on Tori,''} \jhep{02}{1998}{003}, \hepth{9711162}}

\bibitem{DH}{M.R.~Douglas and C.~Hull,
  {``D-branes and the Noncommutative Torus,''}
  \jhep{02}{1998}{008}, \hepth{9711165}} 

\bibitem{SWNCG}{N. Seiberg and E. Witten,
  {``String Theory and Noncommutative Geometry,''}
 \jhep{9909}{1999}{032},\hepth{9908142}}

\bibitem{NekSch}{N.Nekrasov, A.Schwarz,
  {``Instantons on noncommutative $R^4$ and $(2,0)$
  superconformal six dimensional theory,''} \hepth{9802068}}

\bibitem{Berk}{M. Berkooz,
  {``Non-local Field Theories and the Non-commutative Torus,''}
  \hepth{9802069}}

\bibitem{MSV}{S. Minwalla, M. van Ramsdonk and N. Seiberg,
  {``Noncommutative Perturbative Dynamics,''} 
  \hepth{9912072}}

\bibitem{MST}{A. Matusis, L. Susskind and N. Toumbas,
  {``The IR/UV Connection in the Noncommutative Gauge Theories,''}
  \hepth{0002075}}

\bibitem{Arm}{A. Armoni,
  {``Comments on Perturbative Dynamics of Noncommutative Yang-Mills
  Theory,''} \hepth{0005208}}

\bibitem{KKO}{A. Kapustin, A. Kuznetsov, D. Orlov,
  {``Noncommutative Instantons and Twistor Transform,''}
  \hepth{0002193}}

\bibitem{FMW}{R. Friedman, J. Morgan, and E. Witten,
  {``Vector Bundles And F Theory,''}
  \cmp{187}{1997}{679}, \hepth{9701162}}

\bibitem{BJPS}{M. Bershadsky, A. Johansen, T. Pantev and V. Sadov,
  {``Four-Dimensional Compactifications of F-theory,''}
  \np{505}{1997}{165}, \hepth{9701165}}

\bibitem{BFSS}{T. Banks, W. Fischler, S.H. Shenker, L. Susskind,
  {``M-Theory As A Matrix Model: A Conjecture,''} \hepth{9610043}}

\bibitem{ABKSS}{O. Aharony, M. Berkooz, S. Kachru,
  N. Seiberg, and E. Silverstein,
  {``Matrix Description of Interacting Theories in Six Dimensions,''}
  \atmp{1}{1998}{148}, \hepth{9707079}}

\bibitem{WitQHB}{E. Witten,
  {``On The Conformal Field Theory Of The Higgs Branch,''}
  \hepth{9707093}}

\bibitem{GanSet}{O.J. Ganor and S. Sethi,
  {``New Perspectives On Yang-Mills Theories
  with 16 Supersymmetries,''}
     \jhep{01}{1998}{007}, \hepth{9712071}}

\bibitem{ABS}{O. Aharony, M. Berkooz and N. Seiberg,
  {``Light-Cone Description of $(2,0)$ Superconformal
  Theories in Six Dimensions,''} \hepth{9712117}} 

\bibitem{IntSei}{K. Intriligator and N. Seiberg,
  {``Mirror Symmetry in Three-Dimensional Gauge Theories,''}
  \hepth{9607207}}

\bibitem{SeiIRD}{N. Seiberg,
  {``IR Dynamics on Branes and Space-Time Geometry,''}
  \pl{384}{1996}{81--85}, \hepth{9606017}}

\bibitem{KacVaf}{S. Kachru and C. Vafa,
  {``Exact Results For $N=2$ Compactifications
  Of Heterotic Strings,''} \np{450}{95}{69}, \hepth{9505105}} 

\bibitem{DouMoo}{Michael R. Douglas and G. Moore, 
         {``D-branes, Quivers, and ALE Instantons,''}
         \hepth{9603167}}

\bibitem{SWThreeD}{N. Seiberg and E. Witten,
  {``Gauge Dynamics And Compactification To Three Dimensions,''}
  \hepth{9607163}}

\bibitem{IntNEW}{K. Intriligator,
  {``New String Theories in Six Dimensions via Branes at Orbifold
  Singularities,''} \hepth{9708117}}

\bibitem{CGKM}{Y.-K.E.~Cheung, O.J.~Ganor,
  M.~Krogh and A.Yu.~Mikhailov, {``Noncommutative Instantons and
 Twisted (2,0) and Little String Theories,''} \hepth{9812172}}

\bibitem{KapSet}{A. Kapustin and S. Sethi,
  {``The Higgs Branch of Impurity Theories,''}
  \hepth{9804027}}

\bibitem{SeiVBR}{N. Seiberg,
  {``Matrix Description of M-theory on $T^5$ and $T^5/Z_2$,''}
  \pl{408}{97}{98}, \hepth{9705221}}

\bibitem{IntCOM}{K. Intriligator,
  {``Compactified Little-String Theories and Compact
  Moduli Spaces of Vacua,''}
  \hepth{9909219}}

\bibitem{SWYM}{N. Seiberg and E. Witten,
  {\em ``electric - magnetic duality, monopole condensation,
  and confinement in N=2 supersymmetric yang-mills theory''},
  \np{426}{94}{19}, \hepth{9407087}} 

\bibitem{AsNiSc}{A.Astashkevich, N. Nekrasov and A. Schwarz,
  {``On Noncommutative Nahm Transform,''} 
  \cmp{211}{2000}{167}, \hepth{9810147}}

\bibitem{ASchw}{A.~Schwarz,
  {``Morita equivalence and duality,''}
  \np{534}{1998}{720},\hepth{9805034}}

\bibitem{RSchw}{M. A. Rieffel and A. Schwarz,
  {``Morita Equivalence of Multidimensional Noncommutative Tori,''}
 math.QA/9803057}

\bibitem{MZ}{B.~Morariu and B.~Zumino,
  {``Super Yang-Mills on the noncommutative torus,''}
  \hepth{9807198}}

\bibitem{BraMor}{D.~Brace and B.~Morariu,
  {``A Note on the BPS Spectrum of the Matrix Model,''}
 \jhep{02}{1999}{004}, \hepth{9810185}}

\bibitem{BMZ1}{D.~Brace and B.~Morariu and B. Zumino,
  {``Dualities Of The Matrix Model {}From $T$ Duality
  Of The Type II String,''} \np{545}{1999}{192},
  \hepth{9810099}}

\bibitem{KonSch}{A. Konechny and A. Schwarz,
  {``BPS states on noncommutative tori and duality,''}
  \np{550}{1999}{561}, \hepth{9811159}}

\bibitem{BMZ2}{D.~Brace and B.~Morariu and B. Zumino,
  {``$T$ Duality And Ramond-Ramond Backgrounds In The Matrix Model,''}
  \np{549}{1999}{181}, \hepth{9811213}}

\bibitem{HV1}{C.~Hofman and E.~Verlinde,
  {``Gauge bundles and Born-Infeld on the noncommutative torus,''}
  \np{547}{1999}{157}, \hepth{9810219}}

\bibitem{HV2}{C.~Hofman and E.~Verlinde,
  {``U-duality of Born-Infeld on the noncommutative two-torus,''}
  \jhep{9812}{1998}{010}, \hepth{9810116}}

\bibitem{RW}{I. Raebrun and D.P. Willimans,
  {\em ``Morita Equivalence and Continuous-Trace $C^{*}$-Algebras,''}
  American Mathematical Society, 1998}

\bibitem{PS}{B.~Pioline and A.~Schwarz,
  {``Morita equivalence and T-duality (or B versus Theta),''}
  \jhep{9908}{1999}{021}, \hepth{9908019}}

\bibitem{SJ}{M.M. Sheikh-Jabbari,
  {``Noncommutative Super Yang-Mills Theories with 8 Supercharges
  and Brane Configurations,''} \hepth{0001089}}

\bibitem{GriHar}{P.Griffiths, J.Harris,
  {\it "Principles of Algebraic Geometry",}John Wiley and Sons, Inc. 1994}

\bibitem{Konts}{M. Kontsevich,
  {``Deformation Quantization of Poisson Manifolds - I,''}
  q-alg/9709040}

\bibitem{Fed}{B.Fedosov,
  {``A Simple Geometrical Construction of Deformation
  Quantization,''} J. Diff. Geom., 40 (1994) 213-238.}

\bibitem{CatFel}{A.S. Cattaneo and G. Felder,
  {``A Path Integral Approach to the Kontsevich 
  Quantization Formula,''} math.QAS/9902090}

\bibitem{WilCom}{M. de Wilde and P.B.A. Le Compte, "Existence of 
  star-products on exact symplectic manifolds", Annales de l'Institut
  Fourier 35(1985) 117-143}

\bibitem{LMST}{W. Lerche, R. Minasian, C. Schweigert and S. Theisen,
 {``A Note on the geometry of CHL heterotic strings,''}
 \pl{424}{1998}{53-59}, \hepth{9711104}}

\bibitem{Bian}{M.Bianchi,
  {``A Note on Toroidal Compactifications of the Type-I Superstring
   and Other Superstring Vacuum Configurations with 16 Supercharges,''}
  \np{528}{1998}{73},\hepth{9711201}}

\bibitem{WitV}{E. Witten,
 {``Compactification without vector structure,''}
 \jhep{9802}{1998}{006}, \hepth{9712028}}

\bibitem{WitSMI}{E. Witten,
  {``Small Instantons in String Theory''},
  \np{460}{96}{541}, \hepth{9511030}}

\bibitem{GMS}{O.J. Ganor, D.R. Morrison and N. Seiberg,
  {``Branes, Calabi-Yau Space, and Toroidal
  Compactification of the $N=1$ Six-Dimensional $E_8$ Theory,''}
  \np{487}(1997){93}, \hepth{9610251}}

\bibitem{GanC}{O.J. Ganor,
  {``Toroidal Compactification of Heterotic 6D
  Non-Critical Strings Down to Four Dimensions,''}
  \np{488}(1997){223}, \hepth{9608109}}

\bibitem{DLS}{M. R. Douglas, D. A. Lowe and J. H. Schwarz,
  {``Probing F-theory With Multiple Branes,''} \hepth{9612062}}

\bibitem{Sch}{J.Schwarz, 
  {``Noncompact Symmetries in String Theory,''} \hepth{9207016}}

\bibitem{DLM}{M.Duff, J.Liu, R.Minasian, 
  {``Eleven dimensional origin of
  string/string duality: a one loop test,''} \hepth{9506126}}

\bibitem{RD}{R. Dijkgraaf, 
  {``Instanton Strings and Hyperkahler geometry,''} \hepth{9810210}}

\bibitem{MK}{S. Mukai, 
 {``On the moduli space of bundles on $K_3$ surfaces I,''}
   in M.F. Atiyah et al Eds., 
  Vector Bundles on Algebraic Varieties, (Oxford, 1987)}

\bibitem{SenKK}{A. Sen,
  {``Dynamics of Multiple Kaluza-Klein Monopoles in M-theory
   and String Theory''} \atmp{115}{1998}{1}, \hepth{9707042}}

\bibitem{WitKK}{E. Witten,
  {``Heterotic String Conformal String Theory and ADE Singularity,''}
  \hepth{9909229}}

\bibitem{Rozali}{M. Rozali,
  {``Hypermultiplet moduli space and three dimensional
   gauge theories,''} \hepth{9910238}}

\bibitem{Krogh}{M. Krogh,
  {``S-Duality and Tensionless 5-branes in Compactified Heterotic
  String Theory,''}
  \jhep{9912}{018}{1999}, \hepth{9911084}}

\bibitem{AspMor}{P. Aspinwall and D. Morrison,
  {``Point-like Instantons on $K_3$ Orbifolds,''} 
  \np{503}{1997}{533}, \hepth{9705104}}

\bibitem{VafID}{C. Vafa,
  {``Instantons on D-branes,''}
  \np{463}{1996}{435}, \hepth{9512078}}

\bibitem{KKKLLY}{
  E. Kim, H. Kim, N. Kim, B.-H. Lee, C.-Y. Lee, H. Seok Yang,
  {``Matrix Theory and D-brane Bound States on Noncommutative $\MT{4}$,''}
  \hepth{9912272}}







\bibitem{WitBND}{E.~Witten,
  {``Bound States of Strings and $p$-Branes,''}
  \np{460}{1996}{335--350}, \hepth{9510135}}

\bibitem{StrOPN}{A. Strominger,
  {``Open $p$-Branes,''}
  \pl{383}{1996}{44-47}, \hepth{9512059}}

\bibitem{BDS}{T. Banks, M.R. Douglas and N. Seiberg,
  {``Probing F-theory With Branes,''}
  \pl{387}{1996}{278--281}, \hepth{9605199}}

\bibitem{BluInt}{J.D. Blum and K. Intriligator,
  {``New Phases of String Theory and 6d RG Fixed Points via Branes at
  Orbifold Singularities,''} \np{506}{1997}{199},\hepth{9705044}}

\bibitem{Connes}{A. Connes, Noncommutative geometry, Academic Press(1994)}

\bibitem{WitBR}{E.~Witten,
  {``Solutions Of Four-Dimensional Field Theories Via M Theory,''}
  \np{500}{1997}{3--42},\hepth{9703166}}

\bibitem{Don}{S.K. Donaldson,
  {``An application of gauge theory to four
  dimensional topology,''} J.Diff.Geom.18, 279.}

\bibitem{Ulen}{K.Uhlenbeck and S.T.Yau, (1986) preprint.}



\bibitem{Andrei}{A.Yu. Mikhailov,
  {``D1-D5 System and Noncommutative Geometry,''}
  \hepth{9910126}}

\bibitem{Vipul}{V. Periwal,
  {``Nonperturbative effects in Deformation Quantization,''}
  \hepth{0006001}}

\end{thebibliography}
\end{document}